\documentclass[10pt,a4paper,openright,tablecaptionabove]{scrartcl}
\usepackage{subfig}
\label{key}
\usepackage[english]{babel}
\usepackage{amsmath, amsthm, amssymb}
\usepackage{bbm}
\usepackage{url}
\usepackage{ulem}
\usepackage{tikz-cd}
\usepackage{braket}
\usepackage{path}
\usepackage{nicefrac}
\usepackage{amssymb}
 \usetikzlibrary{shapes,arrows}
\usepackage{amsmath}
\usepackage{amsthm}
\usepackage{cancel}
\usepackage{indentfirst}
\usepackage{eucal}
\usepackage{appendix}
\usepackage[utf8]{inputenc}
\usepackage{geometry}
\usepackage{verbatim}
\usepackage{nameref}
\usepackage{alltt}
\usepackage{hyperref}
\usepackage{pgfplots}
\usepackage{dsfont}
\usepackage{appendix}
\usepackage[font=small, labelfont=bf]{caption} 

\newtheorem{Definition}{Definition}

\tikzstyle{decision} = [diamond, draw, fill=white!20, 
text width=4.5em, text badly centered, node distance=3cm, inner sep=0pt]
\tikzstyle{block} = [rectangle, draw, fill=white!20, 
text width=4.5em, text centered, rounded corners, minimum width=4em, minimum height=4em]
\tikzstyle{line} = [draw, -latex']
\tikzstyle{cloud} = [draw, circle, fill=white!20, node distance=3cm,
minimum height=2em]

\areaset[current]{450pt}{720pt}

\numberwithin{equation}{section}

\begin{document}

\thispagestyle{empty}

\vskip 0.5cm

\thispagestyle{empty}
\begin{center}
\Large{{\bf Algebraic (super-)integrability from commutants of subalgebras in universal enveloping algebras }}
\end{center}
\vskip 0.5cm
\begin{center}
\small{\textsc{Rutwig Campoamor-Stursberg$^{1,\star}$, Danilo Latini$^{2,*}$, Ian Marquette$^{2,\dagger}$ and Yao-Zhong Zhang$^{2,\ddagger}$}}
\end{center}
\begin{center}
\small{$^1$ Instituto de Matem\'{a}tica Interdisciplinar and Dpto. Geometr\'{i}a y Topolog\'{i}a, UCM, E-28040 Madrid, Spain}
\end{center}
\begin{center}
\small{$^2$ School of Mathematics and Physics, The University of Queensland, Brisbane, QLD 4072, Australia}
\end{center}
\begin{center}
	\footnotesize{$^\star$\textsf{rutwig@ucm.es} \hskip 0.25cm$^*$\textsf{d.latini@uq.edu.au} \hskip 0.25cm $^\dagger$\textsf{i.marquette@uq.edu.au} \hskip 0.25cm $^\ddagger$\textsf{yzz@maths.uq.edu.au}}
\end{center}
\vskip  1cm
\hrule
\begin{center}
	{\bf Abstract}
\end{center}
\begin{abstract}
 \noindent  Starting from a purely algebraic procedure based on the commutant of a subalgebra in the universal enveloping algebra of a given Lie algebra, the notion of algebraic Hamiltonians and the constants of the motion generating a polynomial symmetry algebra is proposed.  The case of the special linear Lie algebra $\mathfrak{sl}(n)$ is discussed in detail, where an explicit basis for the commutant with respect to the Cartan subalgebra is obtained, and the order of the polynomial algebra is computed. It is further shown that, with an appropriate realization of $\mathfrak{sl}(n)$, this provides an explicit connection with the generic superintegrable model on the $(n-1)$-dimensional sphere $\mathbb{S}^{n-1}$ and the related Racah algebra $R(n)$.  In particular, we  show explicitly how the models on the $2$-sphere and $3$-sphere and the associated symmetry algebras can be obtained  from the quadratic and cubic polynomial algebras generated by the commutants defined in the enveloping algebra of $\mathfrak{sl}(3)$ and $\mathfrak{sl}(4)$, respectively. The construction is performed in the classical (or Poisson-Lie) context, where the Berezin bracket replaces the commutator. \newline
 \vskip 0.1cm
 \end{abstract}
\vskip 0.35cm
\hrule

\section{Introduction}

\noindent A lot of work has been devoted in recent years to a full characterization and structural comprehension of exactly solvable, integrable and superintegrable systems on various classes of spaces, both at the classical and quantum level \cite{perel,Tem, Trem,Kalnins_2007,Mill13}. Several new connections with different areas as nonlinear differential equations, the theory of special functions, the Painlevé transcendents and algebraic structures beyond Lie algebras have emerged from this work, providing alternative techniques and tools complementary to the traditional analytical approach. In particular, certain types of algebras have been shown to be of special interest in this context, from which several criteria for the construction of (super-)integrable models have been deduced (see e.g. \cite{Frei,Ball,Das,Jar}). 

\medskip
More recently, it was observed that quadratically superintegrable systems can be connected with purely algebraic schemes, where the integrals of the motion are in fact associated to an underlying Lie algebra and the corresponding universal enveloping algebra \cite{CSIM22a, CSIM21a}. Albeit these approaches are mainly based on specific superintegrable systems realized by differential operators, the ansatz can be extrapolated to arbitrary pairs of Lie algebras and subalgebras without reference to a given realization \cite{Cam2022}, hence allowing us to define generically the notion of algebraic Hamiltonians and their corresponding (algebraic) constants of the motion. 

\medskip
The purpose of this paper is the following. Starting from the conventional notion of a commutant in the enveloping algebra of a Lie algebra $\mathfrak{s}$, we reformulate the problem by means of the Lie-Poisson structure of the corresponding symmetric algebra $S(\mathfrak{s})$, which provides a computationally more adequate frame. We then define the notion of algebraic Hamiltonian with respect to a subalgebra $\mathfrak{a}$, and show that the commutant defines a polynomial algebra that can be identified with the symmetry algebra of the Hamiltonian, with the constants of the motion obtained from the elements belonging to the centralizer $C_{S(\frak{s})}\left( \frak{a}\right)$. As an illustrating example of the procedure, we analyze the commutant of the Cartan subalgebra $\mathfrak{h}$ in the special linear algebra $\mathfrak{sl}(n)$ and obtain explicit expressions for the linearly independent elements in the enveloping algebra. It is further shown that the polynomial algebra $\mathcal{A}_n$ generated by these linearly independent monomials is of order $n-1$ for $n\geq 3$. For values $n \geq 3$, the polynomial algebras $\mathcal{A}_n$ are associated, via an appropriate realization of $\mathfrak{sl}(n)$, with the Racah algebra $R(n)$, corresponding to the symmetry algebra of the generic superintegrable models on the sphere $\mathbb{S}^{n-1}$ (see \cite{bie2018racah, bie2020racah, gab19,Correa2020,Lati21} and references therein). The reduction of $\mathcal{A}_n$ to $R(n)$ is explicitly computed for $n=3,4$, with the general case outlined due to high dimension of the polynomial algebras.

\section{The commutant of Lie subalgebras in enveloping algebras}

\noindent Let $\mathfrak{s}$ be an $n$-dimensional real or complex Lie algebra and $ \mathcal{ U}(\mathfrak{s})$ be its universal enveloping algebra. For positive integers $p$, we define $\mathcal{U}_{(p)}(\mathfrak{s})$ as the subspace generated by monomials $X_1^{a_1}\dots X_n^{a_n}$ subjected to the constraint $a_1+a_2+\dots +a_n\leq p$, where $\{X_1,\dots ,X_n\}$ is an arbitrary basis of $\mathfrak{s}$. In this context, an element $P\in \mathcal{U}(\mathfrak{s})$ has degree $d$ if  $d={\rm inf}\left\{k\;|\; P\in \mathcal{U}_{(k)}(\mathfrak{s})\right\}$.  Due to the natural filtration in $ \mathcal{ U}(\mathfrak{s})$, for any $p,q\geq 0$ we have 
\begin{equation}
\mathcal{U}_{(0)}(\mathfrak{s})=\mathbb{C},\; \mathcal{U}_{(p)}(\mathfrak{s})\mathcal{U}_{(q)}(\mathfrak{s})\subset \mathcal{U}_{(p
+q)}(\mathfrak{s}). \label{fil1}
\end{equation}
It follows in particular that any $\mathcal{U}_{(p)}(\mathfrak{s})$ is a finite-dimensional representation of $\mathfrak{s}$, a fact that allows us to represent $\mathcal{U}(\mathfrak{s})$ as a sum of finite-dimensional representations of $\mathfrak{s}$ (see \cite{Dix}). 

\medskip
\noindent In terms of a given basis, the adjoint action of $\mathfrak{s}$ on $\mathcal{U}(\mathfrak{s})$ and on the associated symmetric algebra $S(\mathfrak{s})$ is respectively given by 
\begin{equation}\label{adja}
\begin{array}[c]{rl}
P\in \mathcal{U}(\mathfrak{s}) \mapsto & P.X_i:= \left[X_i,P\right]= X_i P-P X_i\in\mathcal{U}(\mathfrak{s}),\\
P\left(x_1,\dots ,x_n\right)\in S(\mathfrak{s})\mapsto &\displaystyle  \widehat{X}_i(P)=C_{ij}^{k} x_k \frac{\partial P}{\partial x_j}\in S(\mathfrak{s}),\\
\end{array},
\end{equation}
where in particular the differential operators $\widehat{X}_i=C_{ij}^{k} x_k \displaystyle\frac{\partial}{\partial x_j}$ correspond to infinitesimal generator of the one-parameter subgroup associated to the generators $X_i$ by the coadjoint representation \cite{Ra}. The symmetric algebra $S\left( \frak{s}\right) $, that we can identify with $%
\mathbb{K}\left[ x_{1},\ldots ,x_{n}\right] $ with $\mathbb{K}=\mathbb{R,C}$, admits naturally a structure
of Poisson algebra through the prescription\footnote{Usually called the Lie-Poisson or Berezin bracket \cite{Ber}.}  
\begin{equation}
\left\{ P,Q\right\} =C_{ij}^{k}x_{k}\frac{\partial P}{\partial x_{i}}\frac{%
\partial Q}{\partial x_{j}},\;P,Q\in S\left( \frak{s}\right) .  \label{Ber}
\end{equation}
It follows that $\left( S\left( \frak{s}\right) ,\left\{ ,\right\} \right) $ is a Lie algebra containing a subalgebra isomorphic to $\frak{s}$. By means of the symmetrization map
\begin{equation}\label{syma}
\Lambda\left(x_{j_1}\dots x_{j_p}\right)=\frac{1}{p!} \sum_{\sigma\in  \Sigma_{p}} X_{j_{\sigma(1)}}\dots X_{j_{\sigma(p)}},
\end{equation}
with $\Sigma_p$ being the symmetric group of $p$ letters, we obtain the canonical linear isomorphism $\Lambda:S(\mathfrak{s})\rightarrow \mathcal{U}(\mathfrak{s})$ that commutes with the adjoint action. If $S^{(p)}(\mathfrak{s})$ denotes the homogeneous polynomials of degree $p$, the identity $\mathcal{U}^{(p)}(\mathfrak{s})=\Lambda\left(S^{(p)}(\mathfrak{s})\right)$ induces the decomposition $\mathcal{U}_{(p)}(\mathfrak{s})=\sum_{k=0}^{p} \mathcal{U}^{(k)}(\mathfrak{s})$, implying that for arbitrary $P\in \mathcal{U}_{(p)}(\mathfrak{s}), Q\in\mathcal{U}_{(q)}(\mathfrak{s})$, the relation 
\begin{equation*}
\left[ P,Q\right]\in \mathcal{U}_{(p+q-1)}(\mathfrak{s})
\end{equation*}
holds. A distinguished (Abelian) subalgebra is given by the centre of $\mathcal{U}(\mathfrak{s})$
\begin{equation}\label{INVS1}
Z\left(\mathcal{U}(\mathfrak{s})\right) = \left\{ P\in\mathcal{U}(\mathfrak{s})\; |\;\left[\mathfrak{s},P\right]=0\right\},
\end{equation}
consisting of the invariant polynomials of $\mathfrak{s}$. Within the commutative frame with the Lie-Poisson bracket, we have the centre 
\begin{equation*}
Z\left(S\left( \frak{s}\right)\right)=\left\{ P\in S\left( \frak{s}\right) \;|\;\left\{ P,Q\right\} =0,\;Q\in
S\left( \frak{s}\right) \right\} ,
\end{equation*}
which is not only linearly but also algebraically isomorphic to $Z\left(\mathcal{U}(\mathfrak{s})\right)$, albeit only for the class of nilpotent Lie algebras this algebraic isomorphism coincides with $\Lambda$ (see e.g. \cite{Dix59}).

\begin{Definition} 
The commutant $C_{\mathcal{U}(\mathfrak{s})}(\mathfrak{a})$ of a subalgebra $\mathfrak{a}\subset \mathfrak{s}$ is defined as the centralizer of $\mathfrak{a}$ in $\mathcal{U}(\mathfrak{s})$ : 
\begin{equation}
C_{\mathcal{U}(\mathfrak{s})}(\mathfrak{a})=\left\{ Q\in\mathcal{U}(\mathfrak{s})\; |\; [P,Q]=0,\quad \forall P\in\mathfrak{a}\right\}.\label{comm}
\end{equation}
\end{Definition}
Commutants of subalgebras can have quite a complicated structure, and no generic characterization exists for arbitrary types of Lie algebras and subalgebras. However, for semisimple or reductive Lie algebras, there are several criteria that allow their study (see e.g. \cite{Dix}). In this context, it can be ensured that the commutant $C_{\mathcal{U}(\mathfrak{s})}(\mathfrak{a})$ is Noetherian and finitely generated whenever the subalgebra $\mathfrak{a}$ is reductive in $\mathfrak{s}$\footnote{This means that for any element $X$ in $\mathfrak{a}$ the adjoint operator is semisimple \cite{Dix}.}. We call $\left\{P_1,\dots ,P_s\right\}$ a (linear) basis of the commutant $C_{\mathcal{U}(\mathfrak{s})}(\mathfrak{a})$ if it is spanned as vector space by the elements  
\begin{equation}\label{bas1}
P_1^{a_1}P_2^{a_2}\dots P_s^{a_s},\quad a_i\in\mathbb{N}\cup {0},
\end{equation}
and where the coefficients $a_i$ are subjected to some algebraic constraints. It is important to note that these polynomials are linearly but generally not algebraically independent. The requirement on linear independence is necessary if the elements in $C_{\mathcal{U}(\mathfrak{s})}(\mathfrak{a})$ are supposed to generate a (finite-dimensional) polynomial algebra. If $P,Q\in C_{\mathcal{U}(\mathfrak{s})}(\mathfrak{a})$, the Jacobi identity implies that  
\begin{equation}
\left[\mathfrak{a},[P,Q]\right]+\left[Q,[\mathfrak{a},P]\right]+\left[P, [Q,\mathfrak{a}]\right]=0, 
\end{equation}
and since $PQ$ annihilates the subalgebra $\mathfrak{a}$, it admits an expression of the type (\ref{bas1}) and is thus specified by at most $s$ scalars $a_i$ (see (\ref{bas1})). In this sense, we say that the linear dimension of the commutant is $s$, denoting it by $\dim_{L} C_{\mathcal{U}(\mathfrak{s})}(\mathfrak{a})=s$. As a general rule, for a maximal set of algebraically independent polynomials that commute with $\mathfrak{a}$, we do not obtain a polynomial algebra with respect to the commutator, as the algebraic dependence relations may be determined by rational non-polynomial functions, hence not belonging to the enveloping algebra, but to its field of fractions \cite{Alo}.   

\medskip
A commutant $C_{\mathcal{U}(\mathfrak{s})}(\mathfrak{a})$ contains, in particular, the invariant polynomials of the subalgebra $\mathfrak{a}$, in addition to the Casimir operators (whenever these exist) $\mathcal{C}_1,\dots , \mathcal{C}_\ell$ of $\mathfrak{s}$. With respect to the latter,  the commutant $C_{\mathcal{U}(\mathfrak{s})}(\mathfrak{a})$ has the structure of a free module over $\mathbb{C}\left[\mathcal{C}_1,\dots , \mathcal{C}_\ell\right]$, as follows at once from the Schur lemma (see \cite{Dix}). 

\medskip
 Using the canonical isomorphism $\Lambda$  allows us to translate the problem of
determining the commutant of a subalgebra $\frak{a}$ of $\frak{s}$ in the
enveloping algebra $\mathcal{U}\left( \frak{s}\right) $ to the Lie-Poisson
context: for $\frak{a}^{\ast }\subset \frak{s}^{\ast }$ we define the
centralizer 
\begin{equation*}
C_{S(\frak{s})}\left( \frak{a}\right) =\left\{ Q\in S\left( 
\frak{s}\right) \;|\;\left\{ P,Q\right\} =0,\;P\in \frak{a}\right\} .
\end{equation*}
Elements in the latter space are obtained as (polynomial) solutions of the
system of partial differential equations 
\begin{equation}
\widehat{X_{i}}\left( Q\right) :=\left\{ x_{i},Q\right\} =C_{ij}^{k}x_{k}%
\frac{\partial Q}{\partial x_{j}}=0,\;1\leq i\leq m=\dim \frak{a},
\label{SIS}
\end{equation}
where $\left\{ x_{1},\ldots ,x_{m}\right\} $ are coordinates in a dual basis
of $\frak{a}^{\ast }$ (see equation (\ref{adja})). The number of functionally independent 
solutions of system (\ref{SIS}) is given by $r_0=\dim\mathfrak{s}-{\rm rank}(A)$, where $A$ is the $m\times n$-matrix 
with entries   $\left(C_{ij}^{k}x_{k}\right)$ \cite{Bel66,RM}. As solutions of (\ref{SIS}) are not necessarily polynomials, the number $\xi_0$ of independent polynomials commuting with $\mathfrak{a}$ is upper bounded by $\xi_0\leq r_0$, thus providing a lower bound for the linear dimension of the 
commutant: $\dim_{L}   C_{\mathcal{U}(\mathfrak{s})}(\mathfrak{a})\geq \xi_0$. 
 
\subsection{Algebraic integrability and superintegrability}

The analysis of superintegrable systems from the perspective of algebraic structures (see e.g. \cite{Das,gen14,Jar, Correa2020, lat21}) suggest to consider a more general frame, which is suitably placed within enveloping algebras. In this context, various approaches have recently been proposed to define an algebraic notion of integrability and superintegrability \cite{CSIM22a,Cam2022}.  

\begin{Definition}
Let $\mathfrak{a}\subset \mathfrak{s}$ be a Lie subalgebra and $C_{\mathcal{U}(\mathfrak{s})}(\mathfrak{a})$ be the commutant. An algebraic Hamiltonian 
with respect to $\mathfrak{a}$ is defined as  
\begin{equation}\label{hamfu}
\mathcal{H}_a= \sum_{i,j} \alpha_{ij} X_iX_j + \sum_k \beta_k X_k + \sum_{\ell} \gamma_{\ell} \mathcal{C}_{\ell},
\end{equation}
where $X_i,X_j,X_k \in \mathfrak{a}$, $\mathcal{C}_{\ell}$ is a Casimir operator of $\mathfrak{s}$ and $\alpha_{ij}$, $\beta_k$, $\gamma_{\ell}$ are parameters. 
\end{Definition}
The advantage of such a definition relies upon the fact that the commutant automatically provides a set of constants of the motion. If $P\in C_{\mathcal{U}(\mathfrak{s})}(\mathfrak{a})$, it commutes with both the subalgebra and the Casimir operators of $\mathfrak{s}$, hence satisfies the condition $\left[\mathcal{H}_a,P\right]=0$. The notion can be easily translated to the Lie-Poisson frame, considering the Hamiltonian 
\begin{equation}\label{hamfu}
\mathcal{H}= \sum_{i,j} \alpha_{ij} x_ix_j + \sum_k \beta_k x_k + \sum_{\ell} \gamma_{\ell} c_{\ell},
\end{equation}
with $c_{\ell}$ being the symmetric counterpart of the Casimir operators. If $P=\lambda(P_0)$ for some $P_0\in C_{S(\frak{s})}\left( \frak{a}\right)$, then clearly $\{\mathcal{H},P_0\}=0$. Depending on the number of functionally independent solutions of system (\ref{SIS}), a sufficient number of constants of the motion that implies the (super-)integrability of the corresponding system can be extracted \cite{CSIM22a,Cam2022}. 

\medskip
The preceding prescription leads to two possible cases that are conveniently separated: 
\begin{enumerate}

\item The polynomials in $C_{S(\frak{s})}\left( \frak{a}\right)$ have all vanishing commutator/Berezin bracket, in which case the algebraic Hamiltonian $\mathcal{H}_a$ is integrable and the symmetry algebra is Abelian.

\item There are non-commuting elements in $C_{S(\frak{s})}\left( \frak{a}\right)$. In this case we say that the algebraic Hamiltonian $\mathcal{H}_a$ possesses a non-commutative (super-)integrability property and the symmetry algebra is a non-Abelian polynomial algebra. The dimension of the symmetry algebra will be given by the number of linearly independent elements among the constants of the motion. 
\end{enumerate}

Once a specific realization (by differential operators) of the Lie algebra $\mathfrak{s}$ has been chosen, the Casimir operators and the previously computed constants of the motion can either decompose or satisfy additional dependence relations in the functional space associated to the realization, imposing further constraints among the generators that can  reduce the dimension of the symmetry algebra. The realized Hamiltonian and invariants eventually still lead to well-defined exactly, quasi-exactly, integrable or even superintegrable systems in the usual sense, i.e. where the variables are in terms of generalized coordinates and their momentum (see e.g. \cite{Frei,Ovs,Vin,Trem,gab19} and references therein). Such an approach has been successfully applied to various Lie algebras such as $\mathfrak{su}(3)$ and $\mathfrak{gl}(3)$ in order to recover the Smorodinsky-Winternitz systems, as well as the generic model on the sphere $\mathbb{S}^2$, which connects to the $58$ superintegrable systems on conformally flat spaces \cite{Mil13,Que}. The approach chosen here is quite different, and is mainly based  on the algebraic setting, where integrals are polynomials in the enveloping algebra of a Lie algebra and the (non-commutative) integrability or superintegrability will be deduced from commutation relations of the polynomials in enveloping algebra of $\mathfrak{s}$, or either using the simpler relations in terms of the Lie-Poisson or Berezin bracket.

\medskip
In the following, the construction will be performed in the classical (or Poisson-Lie) context, where the Berezin bracket replaces the commutator, for computational reasons. It is understood that all the algebraic structures considered in this paper also have a quantum analog, i.e., when using the commutator and the symmetrization map (\ref{syma}).

\section{The commutant of $\mathfrak{h}$ in $\mathfrak{sl}(n)$}
 
 \noindent In this section we compute the commutant of the special linear Lie algebra $\mathfrak{sl}(n)$ with respect to the Cartan subalgebra $\mathfrak{h}$. It is shown that the elements in a basis of polynomials commuting with $\mathfrak{h}$ can be identified with cycles in the symmetric group $\Sigma_n$, thus providing the dimension for arbitrary values of $n$. It is further shown that the resulting polynomial algebras are deeply related, via an appropriate realization, with the description of superintegrable models on the spheres and the Racah algebras \cite{Kalnins_2007,gab19,Lati21}. 

\medskip
\noindent We consider the special linear Lie algebra $\mathfrak{sl}(n)$ in its defining representation. A basis is given by the generators $E_{ij}$ with $1\leq i,j\leq n$ subjected to the constraint $\sum_{i=1}^n E_{ii}=0$. The commutator is then given by 
\begin{equation}\label{kom1}
[E_{ij},E_{kl}]=\delta_{jk} E_{i l}-\delta_{l i} E_{kj} \qquad (1\leq i,j,k,l \leq n) 	 .
\end{equation}
It can be easily verified that a  minimal set of generators for $\mathfrak{sl}(n)$ is given by the $2(n-1)$ elements $E_{i,i+1},E_{i+1,i}$ for $1\leq i\leq n-1$, with the Cartan subalgebra $\mathcal{h}$ being determined by  
\begin{equation}\label{kom1}
[E_{i,i+1},E_{i+1,i}]=E_{i,i}-E_{i+1,i+1}:=H_i \qquad (1\leq i \leq n-1) 	 .
\end{equation}
As the restriction of the Killing form $\kappa$ to the Cartan subalgebra $\mathfrak{h}$ is non-degenerate, $\mathfrak{h}$ is reductive in $\mathfrak{sl}(n)$ and thus the commutant $C_{\mathcal{U}(\mathfrak{sl}(n))}(\mathfrak{h})$ is finitely generated, so that it admits a basis of type (\ref{bas1}). 

\medskip
As already mentioned, for computational purposes, it is more convenient to use the Poisson--Lie setting, i.e., to determine the centralizer $C_{S(\mathfrak{sl}(n))}(\mathfrak{h})$ of $\mathfrak{h}$ in $S(\mathfrak{sl}(n))$. To this extent, we have to solve the system of PDEs 
\begin{equation}
\{f, h_i\}=0,\quad 1\leq i\leq n-1.
\label{CAR}
\end{equation}
Using the analytical approach, it can be easily verified that the system (\ref{CAR}) possesses $\dim\mathfrak{sl}(n)-n+1=n(n-1)$ functionally independent solutions, that can always be chosen as polynomials since the Lie algebra is semisimple. In particular, the Cartan subalgebra $\mathfrak{h}$ is contained in $C_{S(\mathfrak{sl}(n))}(\mathfrak{h})$, and corresponds to the only linear elements in the centralizer. The symmetric counterpart $c^{[2]},\dots, c^{[n]}$ of the Casimir operators of $\mathfrak{sl}(n)$ also belong to $C_{S(\mathfrak{sl}(n))}(\mathfrak{h})$. In order to determine the generic shape of polynomials satisfying the system (\ref{CAR}), suppose that $P=P_1+\dots +P_r$ is a polynomial with homogeneous components $P_k$ of degree $k=1,\dots ,r$. Then $\left\{h_i, P\right\}=\left\{h_i, P_1\right\}+\dots +\left\{h_i, P_r\right\}=0$ if and only if 
\begin{equation}\label{br1}
\left\{h_i, P_k\right\}=0,\quad 1\leq i\leq n-1,\quad 1\leq k\leq r.
\end{equation}
This allows to reduce the analysis of the centralizer to homogeneous polynomials. On the other hand, if $P_k= \lambda^{i_1\dots i_k}x_{i_1}\dots x_{i_k}$ is homogeneous of degree $k$, with $x_{i_j}$ arbitrary elements of $\mathfrak{sl}(n)^{\ast}$, we conclude from the properties of the Lie-Poisson bracket that 
\begin{equation}\label{br2}
\left\{h_\ell, P_k\right\}= \lambda^{i_1\dots i_k}\left(\mu_{i_1}^\ell+\dots +\mu_{i_k}^\ell\right) x_{i_1}\dots x_{i_k},
\end{equation}
where $\mu_{i_j}^\ell$ is the eigenvalue of $x_{i_j}$ with respect to $h_\ell$. As the monomials $x_{i_1}\dots x_{i_k}$ are independent, equation 
(\ref{br1}) is satisfied if and only if  $\mu_{i_1}^\ell+\dots +\mu_{i_k}^\ell=0$ for each $\ell$ and $i_1,\dots ,i_k$. We conclude that it suffices to determine, for each order $d\geq 2$, a maximal set of linearly independent monomials that satisfy the system (\ref{CAR}). Any other element is obtained as a polynomial in these elements. 

\medskip
\noindent For computational purposes, it is convenient to use the following (lexicographically ordered) basis in $\mathfrak{sl}(n)^{\ast}$:
\begin{equation}\label{basisA}
\begin{split}
h_i, & \quad 1\leq i\leq n-1\\
e_{i,i+s},& \quad 1\leq i\leq n-1,\; 1\leq s\leq n-i \\
e_{i+s,i},& \quad 1\leq i\leq n-1,\; 1\leq s\leq n-i.  
\end{split}
\end{equation} 

It follows at once from (\ref{kom1}) that for arbitrary indices $i,j,k$ we have 
\begin{equation}
\left\{h_i, e_{j,k}\right\}=\delta_{i}^{j} e_{i,k}-\delta_{1+i}^{j}e_{i+1,k}-\delta_{i}^{k}e_{j,i}+\delta_{1+i}^{k}e_{j,1+i},
\end{equation}
showing that $\mathfrak{h}^{\ast}$ acts diagonally on the generators of the basis. Rewriting the latter expression as 
\begin{equation}
\left\{h_i, e_{j,k}\right\}=\mu^{i}_{j,k} e_{j,k},
\end{equation}
we denote by $\mu^{i}_{j,k} $ the weight of $e_{j,k}$ with respect to the Cartan generator $h_i$. It can be easily verified that for $1\leq j\leq n-1$ these weights are given by 
\begin{equation}\label{eig1}
\mu^{j-1}_{j,j+1}=-1,\quad  \mu^{j}_{j,j+1}=2,\quad \mu^{j+1}_{j,j+1}=-1,\quad \mu^{\ell}_{j,j+1}=0,\; \ell\neq j-1,j,j+1  .
\end{equation}
If ${\bf A}$ denotes the matrix with entries $\mu^{i}_{j,j+1}$, it follows at once from (\ref{eig1}) that ${\bf A}$ coincides with the Cartan matrix of $\mathfrak{sl}(n)$:
\begin{equation}
{\bf A}=\left( 
\begin{tabular}{ccccccc}
$2$ & $-1$ & $0$ &  &  &  &  \\ 
$-1$ & $2$ & $-1$ &  &  &  &  \\ 
$0$ & $-1$ & $2$ &  &  &  &  \\ 
&  &  & $\ddots $ &  &  &  \\ 
&  &  &  & $2$ & $-1$ & $0$ \\ 
&  &  &  & $-1$ & $2$ & $-1$ \\ 
&  &  &  & $0$ & $-1$ & $2$%
\end{tabular}
\right) 
\end{equation}
This shows that the (column) vectors ${\bf v}_{j,k}=\left(\mu^{1}_{j,k},\dots ,\mu^{n-1}_{j,k}\right)$ with $1\leq j<k\leq n$ actually behave like the positive roots of the root system $\mathcal{R}$ associated to $\mathfrak{sl}(n)$, with the ${\bf v}_{j,j+1}$ corresponding to the simple roots, implying that the ${\bf v}_{j,k}$ are linear combinations of the ${\bf v}_{j,j+1}$ with positive integer coefficients \cite{serr}. Specifically, for any $e_{j,k}$ with $j<k$ we have the following algebraic relation for its weight vector:  
\begin{equation}\label{eig2}
{\bf v}_{j,k}=\sum_{s=j}^{k-1} {\bf v}_{s,s+1},\quad 1\leq j<k\leq n. 
\end{equation}
In analogous way, for the generators $e_{k,j}$ with $j<k$, the eigenvalue vector is given by ${\bf v}_{k,j}=-{\bf v}_{j,k}$. It is worthy to mention two immediate but relevant consequences of (\ref{eig2}):
\begin{enumerate}
\item Linear combinations of the ${\bf v}_{j,k}$ with $j<k$ (respectively $j>k$) and positive integer coefficients cannot be zero, as the vectors ${\bf v}_{j,j+1}$ are linearly independent.

\item For any ${\bf v}_{j,k}$, the multiplicity of the weight vector ${\bf v}_{s,s+1}$ ($1\leq s\leq n-1$) in the sum (\ref{eig2}) is either one or zero.  
\end{enumerate}
These properties enable us to construct recursively a maximal set of linearly independent sets that generate the centralizer. As already mentioned, the only linear solutions to (\ref{CAR})  are the Cartan generators themselves. As any element in $\mathfrak{h}^{\ast}$ has weight zero, it follows that for any monomial $P$ commuting with $\mathfrak{h}^{\ast}$, the product $F(h_1,\dots ,h_{n-1})P$ also satisfies (\ref{CAR}). In order to discard these decomposable solutions, we can suppose without loss of generality that the monomials $P$ satisfying system (\ref{CAR}) additionally fulfill the constraint 
\begin{equation}\label{co1}
\frac{\partial P}{\partial h_\ell}=0,\quad 1\leq \ell\leq n-1.
\end{equation}
We start with $d=2$ and consider quadratic monomials $e_{j,k}e_{l,m}$. By condition (\ref{br2}), the monomial is a solution of system (\ref{CAR}) if and only if the eigenvalue vectors satisfy the constraint ${\bf v}_{j,k}+{\bf v}_{l,m}=0$, hence ${\bf v}_{j,k}=- {\bf v}_{l,m}= {\bf v}_{m,l}$ must hold, implying that $m=j$ and $l=k$. We conclude that the monomial has the form 
\begin{equation}
p_{j,k}= e_{j,k}e_{k,j},\quad 1\leq j<k\leq n.
\end{equation}
As can be easily seen, the number of independent monomials of this type is given by the number of transpositions in the symmetric group $\Sigma_n$, and equals $\nu_2=\frac{n!}{2(n-2)!}$. We observe that the monomials $p_{i,j}$ are not only linearly independent, but moreover functionally independent. This can be easily verified considering the Jacobian matrix $A_J$ with respect to the set of variables $e_{j,k}$ with $j<k$, the determinant of which is given by 
\begin{equation}
\det(A_J)= \prod_{s=1}^{n-1} e_{s+1,s}\neq 0. 
\end{equation}
    
\medskip
Let us now consider $d=3$ and a cubic monomial $e_{i_1,j_1}e_{i_2,j_2}e_{i_3,j_3}$. It is a solution whenever ${\bf v}_{i_1,j_1}+{\bf v}_{i_2,j_2}+{\bf v}_{i_3,j_3}=0$ holds. As we are assuming that the constraint (\ref{co1}) holds, none of the weight vector vanishes. On the other hand, If $i_\alpha < j_\alpha$ for $\alpha=1,2,3$,\footnote{The same holds if  $i_\alpha > j_\alpha$ for $\alpha=1,2,3$.}, the sum of the weight vectors cannot be zero as a consequence of equation (\ref{eig2}). Up to a reordering of the indices, there are two possibilities: either $i_\alpha < j_\alpha$ for $\alpha=1,2$ and $i_3>j_3$ or $i_\alpha > j_\alpha$ for $\alpha=1,2$ and $i_3<j_3$. For both cases, the argumentation is the same, thus let us assume that the first possibility is given. We can further assume that $i_1\leq i_2$. Equation (\ref{eig2}) leads to the identity 
\begin{equation}\label{eig3}
{\bf v}_{i_1,j_1}+{\bf v}_{i_2,j_2}=\sum_{s=i_1}^{j_1-1} {\bf v}_{s,s+1}+\sum_{s=i_2}^{j_2-1} {\bf v}_{s,s+1} =-{\bf v}_{i_3,j_3}={\bf v}_{j_3,i_3}=\sum_{s=j_3}^{i_3-1} {\bf v}_{s,s+1} \, .
\end{equation}
On the right-hand side of the identity, each ${\bf v}_{s,s+1}$ appears with multiplicity one, with the chain from $s=j_3$ to $s=i_3-1$ being uninterrupted. Therefore, the sums of the left-hand side must exhibit the same property. In order to avoid multiplicities, we must have the ordering of indices $i_1<j_1\leq i_2<j_2$, while the identities $i_1=j_3,\; j_1=i_2,\; j_2=i_3$ are a consequence of the fact that the sequence from  $s=j_3$ to $s=i_3-1$ is uninterrupted. We conclude that the cubic monomial is given by 
\begin{equation}\label{3cy}
e_{i_1,j_1}e_{i_2,j_2}e_{i_3,j_3}=e_{i_1,j_1}e_{j_1,j_2}e_{j_2,i_1},
\end{equation}
where the indices $i_1,j_1$ and $j_2$ are all distinct. As before, identifying the index set with the 3-cycle $(i_1j_1j_2)$, we obtain that the number of linearly independent cubic monomials Poisson-commuting with $\mathfrak{h}^{\ast}$ is given by $\nu_3=\frac{n!}{3(n-3)!}$.  

\medskip Before proceeding with the analysis, we observe that the monomial solutions of equation (\ref{CAR}) up to degree three subjected to the constraint (\ref{co1}) contain a maximal set of functionally independent solutions. Considering the monomials $p_{1,j,k}$ with $1<j<k$ jointly with the basis of the Cartan subalgebra and the $p_{r,s}$, we get exactly $n^2-n$ elements, that is, the number of independent solutions of the system (\ref{CAR}). The Cartan generators are clearly independent, so that it suffices to show that the nonlinear polynomials above are independent. We consider the complementary $\Phi$ of the set of variables $\left\{h_1,\dots ,h_{n-1},e_{s,s-1},\; 2\leq s\leq n\right\}$ in the basis (\ref{basisA}), and determine the Jacobian matrix $A_J$ with respect to the variables in $\Phi$. A long but routine computation shows that  
\begin{equation}
\det(A_J)=\prod_{e_{j,k}\notin \Psi} e_{j,k}\times \prod_{k=2}^{n-1}e_{1,k}^{n-k-1}\times \prod_{k=4}^{n}e_{k,1}^{k-3}\neq 0,
\end{equation}
where $\Psi=\left\{e_{1,n},\; e_{k,s},\; 4\leq k\leq n,\; k-2\leq s\leq n-2\right\}$. This shows in particular that the $p_{i,j}$ along with $p_{1,k,l}$ constitute a maximal set of functionally independent solutions of (\ref{CAR}) satisfying the constraint (\ref{co1}). 

\medskip
In order to simplify the analysis of linearly independent monomial solutions $e_{i_1,j_1}\dots e_{i_d,j_d}$ for orders $d\geq4$, we first extract some general consequences from equation (\ref{eig2}). First of all, if there exists some reordering of the indices and an integer $q<d$ such that ${\bf v}_{i_1,j_1}+\dots +{\bf v}_{i_q,j_q}=0$, then clearly 
\begin{equation}
{\bf v}_{i_1,j_1}+\dots +{\bf v}_{i_d,j_d}={\bf v}_{i_{q+1},j_{q+1}}+\dots +{\bf v}_{i_d,j_d}=0
\end{equation}
and the monomial is decomposable as a product of a monomial of order $q$ and a monomial of order $d-q$. We can thus restrict the analysis to those sums of weight vectors for which no partial sum of $q<d$ indices vanishes. We next consider a partition of the index set $\mathcal{S}$ as follows:
\begin{equation}
\mathcal{S}_+=\left\{ i_\alpha < j_\alpha\;|\; 1\leq\alpha\leq d\right\},\quad \mathcal{S}_-=\left\{ i_\alpha > j_\alpha\;|\; 1\leq\alpha\leq d\right\}.
\end{equation}
\medskip
\noindent {\bf Lemma}: If the cardinal ${\rm card}\left(\mathcal{S}_-\right)\geq 2$, then the monomial $e_{i_1,j_1}\dots e_{i_d,j_d}$ is decomposable. 

\medskip
{\bf Proof:} Suppose that ${\rm card}\left(\mathcal{S}_-\right)= 2$. Reordering the indices if necessary, the condition (\ref{br2}) can be rewritten as
\begin{equation}
{\bf v}_{i_1,j_1}+\dots +{\bf v}_{i_{d-2},j_{d-2}}=-{\bf v}_{i_{d-1},j_{d-1}}-{\bf v}_{i_d,j_d}={\bf v}_{j_{d-1},i_{d-1}}+{\bf v}_{j_d,i_d}. 
\end{equation}
Without loss of generality we can suppose that the first indices are ordered lexicographically, i.e., $i_1\leq i_2\leq \dots \leq i_{d-2}$ and $j_{d-1}\leq j_d$. By the decomposition (\ref{eig2})
\begin{equation}
\sum_{s=i_1}^{j_1-1} {\bf v}_{s,s+1} +\dots +\sum_{s=i_{d-2}}^{j_{d-2}-1} {\bf v}_{s,s+1}=\sum_{s=j_{d-1}}^{i_{d-1}-1} {\bf v}_{s,s+1}+\sum_{s=j_{d}}^{i_{d}-1} {\bf v}_{s,s+1} .
\end{equation}
It follows from the sum on the right-hand side that the multiplicity of each ${\bf v}_{s,s+1}$ is at most two. Further, there must exist a partition of the indices such that 
\begin{equation}
\left\{(i_a,j_a),\;\; 1\leq a\leq d-2\right\} = \left\{(\alpha_r,\beta_r),\;\; 1\leq r\leq m_0\right\}\cup \left\{(\lambda_s,\nu_s),\;\; 1\leq s\leq d-2-m_0\right\} 
\end{equation}
and such that 
\begin{equation}
\sum_{s=j_{d-1}}^{i_{d-1}-1} {\bf v}_{s,s+1}=\sum_{r=1}^{m_0} {\bf v}_{\alpha_r,\beta_r},\quad \sum_{s=j_{d}}^{i_{d}-1} {\bf v}_{s,s+1}=\sum_{s=1}^{d-2-m_0} {\bf v}_{\lambda_s,\nu_s} .
\end{equation}
From these identities we conclude that 
\begin{equation}
\begin{split}
j_{d-1}=\alpha_1,\; \beta_r=\alpha_{r+1}\; (1\leq r\leq m_0-1),\; \beta_{m_0}=i_{d-1}, \\
j_d=\lambda_1,\; \nu_s=\lambda_{s+1}\; (1\leq s\leq d-3-m_0),\; \nu_{d-2-m_0}=i_d, 
\end{split}
\end{equation}
showing that the monomial decomposes as 
\begin{equation}
e_{i_1,j_1}\dots e_{i_d,j_d}=\left(\prod_{r=1}^{m_0-1}e_{\alpha_r,\alpha_{r+1}}\right)e_{\alpha_{m_0},\alpha_{1}} 
\left(\prod_{s=1}^{d-2-m_0-1}e_{\lambda_s,\lambda_{s+1}}\right)e_{\lambda_{d-2-m_0},\lambda_1}.  
\end{equation}
The same argument holds for any ${\rm card}\left(\mathcal{S}_-\right)> 2$. $\square$

\medskip
As a consequence of this result, an indecomposable monomial of order $d$ satisfying equations (\ref{CAR}) and (\ref{co1}) satisfies either ${\rm card}\left(\mathcal{S}_-\right)=1$ or ${\rm card}\left(\mathcal{S}_-\right)=d-1$. In the following, we will assume that the first possibility holds, the second being completely equivalent. 
 
\medskip
Let the monomial $e_{i_1,j_1}\dots e_{i_d,j_d}$ satisfy the system (\ref{CAR}). By the preceding lemma, it suffices to consider the case where $i_a<j_a$ for $1\leq a\leq d-1$ and $i_d>j_d$. Writing the condition ${\bf v}_{i_1,j_1}+\dots +{\bf v}_{i_d,j_d}=0$ on the weight vectors as
\begin{equation}
{\bf v}_{i_1,j_1}+\dots +{\bf v}_{i_{d-1},j_{d-1}}=-{\bf v}_{i_d,j_d}={\bf v}_{j_d,i_d} \, ,
\end{equation}
and using again that multiplicities in the decomposition (\ref{eig2}) are at most one, a routine computation shows that the monomial necessarily has the form 
\begin{equation}
p_{i_1,\dots ,i_d}=e_{i_1,i_2}e_{i_2,i_3}\dots e_{i_{d-1},i_d}e_{i_d,i_1}.
\end{equation}
The number of linearly independent elements of this type is given by $\nu_d=\frac{n!}{d(n-d)!}$. It is important to observe that, due to the commutativity of the variables $h_\ell, e_{j,k}$, the cyclic symmetry of indices in the monomials  $p_{i_1,\dots ,i_d}$ gives rise to the same element, i.e. 
\begin{equation}\label{asy}
p_{i_1,\dots ,i_d}= p_{i_2,\dots ,i_d,i_1}=\dots =p_{i_d,i_1\dots ,i_{d-1}}  .
\end{equation}
This property allows us to identify the monomials of degree $d$ satisfying equations (\ref{CAR}) and (\ref{co1}) with the $d-$cycles of $\Sigma_n$. 

\medskip
\noindent {\bf Lemma}: Any monomial $P$ of order $d>n$ satisfying equation (\ref{CAR}) is decomposable. 

\medskip
{\bf Proof} Let $P=e_{i_1,j_1}\dots e_{i_{n+1},j_{n+1}}$ be a monomial solution of (\ref{CAR}). If ${\rm card}\left(\mathcal{S}_-\right)\geq  2$, the previous results show that 
$P$ is decomposable, hence we can suppose that ${\rm card}\left(\mathcal{S}_-\right)=1$. Without loss of generality, we can assume that $i_1\leq i_2\leq \dots \leq i_n$ and that $i_{n+1}>j_{n+1}$. The condition on the eigenvalues reads 
\begin{equation}\label{eig3}
{\bf v}_{i_1,j_1}+\dots +{\bf v}_{i_{n},j_{n}}={\bf v}_{j_{n+1},i_{n+1}}. 
\end{equation}
It is clear that the indices $i_k$ cannot be all distinct, as otherwise, up to a reordering of indices, we would have $i_k=k$ for $1\leq k\leq n$ and thus $k<j_k\leq n$. However, as the right-hand side of (\ref{eig3}) has no multiplicities greater than one, necessarily $j_k=i_{k+1}=(k+1)$ for $1\leq k\leq n-1$ and $j_n=i_{n+1}$. As $i_n=n<j_{n}\leq n$, this leads to a contradiction. Therefore, there must exist at least two equal indices $i_k=i_{k+1}$ for some $1\leq k\leq n-1$. Let $\xi_0={\rm min}(j_k,j_{k+1})$. It follows from the decomposition that all ${\bf v}_{s,s+1}$ with $i_k\leq s\leq \xi_0$ have at least multiplicity two. Again, as each term on the right-hand side of (\ref{eig3}) has multiplicity one, and since by assumption $i_k<j_k$ holds for $k\leq n$, one of the following constraints must hold: 
\begin{equation}
{\bf v}_{i_k,j_k}+\dots +{\bf v}_{i_{\xi_0},j_{\xi_0}}=0\quad {\rm or}\quad {\bf v}_{i_k,j_k}+\dots +{\bf v}_{i_{\xi_1},j_{\xi_1}},
\end{equation}
where $\xi_1={\rm max}(j_k,j_{k+1})$. But each of these possibilities implies that the monomial $P$ is decomposable, proving the assumption. $\square$

\bigskip
Using the analogy with the symmetric group, the result is intuitively clear, as there do not exist $d$-cycles in $\Sigma_n$ for $d>n$.  
Summarizing, it has been shown that a basis of linearly independent elements in the centralizer $C_{S(\mathfrak{sl}(n))}(\mathfrak{h})$ is given by the polynomials 
\begin{equation}\label{bas}
h_\ell,\quad p_{i_1,\dots ,i_d},\quad 1\leq \ell\leq n-1,\quad 2\leq d\leq n.
\end{equation}
Its linear dimension is therefore given by
\begin{equation}\label{dice}
\dim_{L} C_{S(\mathfrak{sl}(n))}(\mathfrak{h}) =(n-1) +\sum_{d=2}^{n}  \nu_d= \sum_{d=1}^{n}\frac{n!}{(n-d)! d}-1  .
\end{equation} 
As only $n^2-n$ of these elements are functionally independent, the basis elements will satisfy several algebraic relations, the number of which increases with the value of $n$. In the following, we present some of these relations. 

\medskip
\noindent
{\bf Proposition:} For any $n\geq 3$ and $3\leq d\leq n$, following dependence relations hold:  
\begin{equation}\label{rela1}
\begin{split}
\left(\prod_{u=1}^{k-1}p_{i_u,i_{u+1}}\right)p_{i_k,i_1}=p_{i_1,i_2,\dots ,i_k}p_{i_1,i_k,i_{k-1},\dots ,i_2}, \\
\prod_{1\leq i<j\leq n}p_{i,j}= p_{1,2,\dots ,n}\; p_{1,n-1,\dots ,2}\prod_{m=1}^{n-2}\prod_{s=m+2}^{n}p_{m,s},\\
\prod_{i_1\neq i_2\neq \dots \neq i_k}p_{i_1,i_2,\dots ,i_k}= \left(\prod_{r<s}p_{r,s}\right)^{\varphi(k)},\quad \varphi(k)=\prod_{s=2}^{k  -1}(n-s). 
\end{split}
\end{equation}
The proof follows by direct verification, taking into account the structure of the monomials (\ref{bas}) and the number (\ref{dice}) of linearly independent elements. 
Although these identities do not exhaust all possible dependence relations, they are useful for simplifying computations. As follows from (\ref{bas}), the commutant $C_{S(\mathfrak{sl}(n))}(\mathfrak{h})$ defines a polynomial algebra $\mathcal{A}_n$ with respect to the Lie-Poisson bracket, and possessing an $(n-1)$-dimensional centre generated by the $h_1,\dots ,h_{n-1}$. 

\medskip
We remark that $n=2$ is the only value where the number (\ref{dice}) of linearly independent monomials coincides with the cardinal of a maximal set of functionally independent solutions of the system (\ref{CAR}). In this case, a basis of the centralizer $C_{\mathfrak{sl}(2)^{\ast}}(\mathfrak{h}^{\ast})$ is given by $h_1,p_{1,2}$, and the resulting algebra is two-dimensional and Abelian. In particular, the symmetric counterpart $c^{[2]}$ of the Casimir operator $\mathcal{C}$ of $\mathfrak{sl}(2)$ is given by $ c^{[2]}=\frac{1}{4}h_1^2+p_{1,2}$.

\bigskip 
For computational purposes, it may be convenient to consider some variant of the basis (\ref{bas}) that takes into account additional symmetry properties, in order to simplify the expression of the Lie-Poisson brackets in the centralizer. For instance, the generic bracket of two quadratic monomials is given by 
\begin{equation}
\begin{split}
\{p_{i_1,j_1}, p_{i_2,j_2}\} = & \quad \delta_{i_1}^{i_2}\left(p_{i_1,j_1,j_2}-p_{i_1,j_2,j_1}\right)+ \delta_{j_1}^{j_2}\left(p_{i_1,i_2,j_2}-p_{i_1,j_2,i_2}\right)\\
 & +\delta_{i_1}^{j_2}\left(p_{i_1,j_1,i_2}-p_{i_1,i_2,j_1}\right)+\delta_{j_1}^{i_2}\left(p_{i_1,j_2,i_2}-p_{i_1,i_2,j_2}\right),
\end{split}
\end{equation}
an identity that suggests to consider an alternative choice of cubic monomials, namely replacing  $p_{i,j,k}$ by their symmetric and skew-symmetric counterparts
\begin{equation}\label{syba}
g_{i,j,k} =\frac{1}{2}\left( p_{i,j,k}+p_{i,k,j}\right),\quad f_{i,j,k} =\frac{1}{2}\left( p_{i,j,k}-p_{i,k,j}\right),
\end{equation} 
and where the coefficient $\frac{1}{2}$ is a consequence of the index cyclic symmetry (\ref{asy}). The same transformation can be considered for higher orders, where it may be  appropriate to consider monomials possessing additional symmetry properties to simplify the expression of the Lie-Poisson bracket. 

\medskip
The polynomial algebra $\mathcal{A}_n$ is of order at most $n-1$ in the generators satisfying the condition (\ref{co1}). This can be easily verified considering for example the monomials of maximal degree $P=p_{1,2,\dots ,n}$ and $Q=p_{1,2,n,n-1,\dots ,4,3}$, where the brackets gives, among other terms 
\begin{equation}
\left\{P,Q\right\}= p_{1,2,3}p_{3,4}\dots p_{n-1,n}p_{2,n}+\dots ,\quad n\geq 4
\end{equation}
and where the first term cannot be written in terms of a lower number of basis elements. This actually corresponds to the maximal possible order of brackets. 
As the bracket of a monomial of degree $d$ (i.e., $2d$ indices) with one of degree $m$ (i.e., $2m$ indices) has degree $d+m-1$, the total number of indices in each terms is given by $2d+2m-2$. Each quadratic monomial $p_{i,j}$ involves four indices, thus the maximal number $\ell_0$ of quadratic terms in a Lie-Poisson bracket is upper bounded by $\ell_0 <\frac{1}{4}(2d+2m-2)$. Longest possible chains can result for $d=m=n$, with $\ell_0 <\frac{1}{4}(4n-2)$. This shows that $\ell_0=n-2$ is the maximal value, with $6$ remaining indices that must correspond to a cubic monomial, hence leading to a maximal length of $n-1$.    

\bigskip Using the canonical chain of embeddings $\mathfrak{sl}(2)\subset \mathfrak{sl}(3)\subset \dots \subset \mathfrak{sl}(n)$, it further follows from the basis (\ref{bas}) that for the polynomial algebra $\mathcal{A}_n$ associated to centralizer $C_{S(\mathfrak{sl}(n))}(\mathfrak{h} )$ we have the ascending filtration $\mathcal{A}_2\subset  \mathcal{A}_3\subset \dots \subset\mathcal{A}_n$, meaning that each $\mathcal{A}_{k}$ can be seen as a non-central extension of $\mathcal{A}_{k-1}$.

\section{$C_{S(\mathfrak{sl}(3))}(\mathfrak{h} )$ and Racah-type algebras}
\noindent For $n=3$, the system (\ref{CAR}) possesses six functionally independent solutions, while the dimension of the centralizer $C_{S(\mathfrak{sl}(3))}(\mathfrak{h} )$, according to formula (\ref{dice}), is seven. Following (\ref{bas}), a basis may be taken as $\mathcal{B}=\left\{h_1,h_2,p_{1,2},p_{1,3},p_{2,3},p_{1,2,3},p_{1,3,2}\right\}$. The algebraic dependence of these elements is given by        
\begin{equation}\label{alde1}
p_{1,2}p_{1,3} p_{2,3}-p_{1,2,3} p_{1,3,2}=0.
\end{equation}
We remark in particular that $p_{1,3,2}$ cannot be expressed as a polynomial in the remaining monomials $ h_1,h_2,p_{1,2},p_{1,3},p_{2,3},p_{1,2,3}$, showing that a maximal set of functionally independent polynomial solutions of (\ref{CAR}) does not generate in general a polynomial algebra. The elements in $\mathcal{B}$ generate a quadratic algebra with nontrivial brackets  
\begin{equation}
\begin{split}
\{p_{1,2}, p_{1,3}\}& =-\{p_{1,2}, p_{2,3}\}=\{p_{1,3}, p_{2,3}\}=p_{1,2,3}-p_{1,3,2},\\
\{p_{1,2}, p_{1,2,3}\}& =p_{1,2}(p_{1,3}-p_{2,3})-h_1 p_{1,2,3},\\
\{p_{1,3}, p_{1,2,3}\}& =p_{1,3}(p_{2,3}-p_{1,2})+(h_1+h_2) p_{1,2,3},\\
\{p_{2,3}, p_{1,2,3}\}& =p_{2,3}(p_{1,2}-p_{1,3})-h_2 p_{1,2,3},\\
\{p_{1,2}, p_{1,3,2}\}& =-p_{1,2}(p_{1,3}-p_{2,3})+h_1 p_{1,3,2},\\
\{p_{1,3}, p_{1,2,3}\}& =-p_{1,3}(p_{2,3}-p_{1,2})-(h_1+h_2) p_{1,3,2},\\
\{p_{2,3}, p_{1,2,3}\}& =-p_{2,3}(p_{1,2}-p_{1,3})+h_2 p_{1,3,2},\\
\{p_{1,2,3}, p_{1,3,2}\}&=h_1 p_{1,3} p_{2,3}+h_2 p_{1,2} p_{1,3}-(h_1+h_2)p_{1,2} p_{2,3}.
\end{split}
\end{equation}
The classical counterpart of the Casimir operators of $\mathfrak{sl}(3)$ turn out to be:
\begin{equation}\label{cas3}
\begin{split}
c^{[2]}=& p_{1,2}+p_{1,3}+p_{2,3}+\frac{1}{3}(h_1^2+h_1 h_2+h_2^2),\\
c^{[3]}= & p_{1,2,3}+p_{1,3,2}+\frac{1}{3}\left((h_1+2 h_2)p_{1,2}+(h_1- h_2)p_{1,3}-(2h_1+ h_2)p_{2,3}\right)+\frac{1}{9}h_1h_2(h_1-h_2)\\
 & +\frac{2}{27}(h_1^3-h_2^3)).
\end{split}
\end{equation}

\subsection{Connection with Racah-type algebras: change of basis}
 
\noindent We show now that there exists an explicit connection of the previous quadratic algebra associated to $C_{S(\mathfrak{sl}(3))}(\mathfrak{h} )$ with a Racah-type quadratic algebra. To this extent, instead of $h_1,h_2$ we consider the three elements 
\begin{equation}
c_1:=\frac{1}{3}(2h_1+h_2) \, , \quad c_2:=\frac{1}{3}(h_2-h_1) \, ,  \quad c_3:=-\frac{1}{3}(h_1+2h_2)  ,
\end{equation}
subjected to the linear relation: 
\begin{equation}
c_1+c_2+c_3=0, \, 
\label{linrel}
\end{equation}
as well as $c_{ij}:=p_{i,j}$. We also introduce skew-symmetric and symmetric elements (see (\ref{syba})) 
\begin{equation}
f_{123}:=\frac{1}{2}(p_{1,3,2}-p_{1,2,3}),\quad g_{123}:=\frac{1}{2}(p_{1,3,2}+p_{1,2,3})  .
\label{eq:newgen}
\end{equation}
In terms of these generators and central elements,  the relations of the quadratic algebra read:
\begin{equation}\label{frel}
\begin{split}
\{c_{12},c_{23}\}& =\{c_{23},c_{13}\}=\{c_{13},c_{12}\}=2\;f_{123},\\
\{c_{12}, f_{123}\}&=(c_{23}-c_{13})c_{12}+(c_1-c_2)g_{123}\\
\{c_{13}, f_{123}\}&=(c_{12}-c_{23})c_{13}+(c_3-c_1)g_{123}\\
\{c_{23}, f_{123}\}&=(c_{13}-c_{12})c_{23}+(c_2-c_3)g_{123} \\
\{c_{12}, g_{123}\}&=(c_1-c_2)f_{123}\\
\{c_{13}, g_{123}\}&=(c_3-c_1)f_{123}\\
\{c_{23}, g_{123}\}&=(c_2-c_3)f_{123}\\
\{f_{123},g_{123}\}&=\frac{1}{2}\bigl((c_1-c_3)c_{12}c_{23}+(c_3-c_2)c_{12}c_{13}+(c_2-c_1)c_{13}c_{23}\bigl), \, 
\end{split}
\end{equation}
to which the algebraic dependency relation 
\begin{equation}
g^2_{123}-f^2_{123}-c_{12}c_{23}c_{13}=0 
\label{funrel}
\end{equation}
must be added. 
The Casimir of the quadratic algebra reads
\begin{equation}
K=f_{123}^2 +c_{12}c_{23}c_{13}-(c_1 c_{23}+c_2 c_{13}+c_3 c_{12})\left(g_{123}-\frac{1}{4}(c_1 c_{23}+c_2 c_{13}+c_3 c_{12})\right)\, .
\label{cas}
\end{equation}
This can be re-expressed in terms of the third order Casimir of $\mathfrak{sl}(3)^{\ast}$ and the central elements as:
\begin{equation}
K=\frac{1}{4}\bigl(c^{[3]}-c_1 c_2 c_3\bigl)^2 \, .
\label{casc}
\end{equation}
\noindent At this point, to make the connection with Racah-type algebras more explicit, we consider the second and third-order Casimir elements $\{c^{[2]}, c^{[3]}\}$ in \eqref{cas3}. The generators are related to the Casimirs through the following functional relations:
\begin{align}
c_{12}+c_{13}+c_{23}&=c^{[2]}-\frac{1}{2}(c_1^2+c_2^2+c_3^2) \label{c2}\\
2 g_{123}- c_3 c_{12}- c_2 c_{13}-c_1 c_{23}&=c^{[3]}-\frac{1}{3}(c_1^3+c_2^3+c_3^3)	\, .
\label{c3}
\end{align}
\noindent Taking into account these relations, and considering the following redefinitions:
\begin{equation}
\bar{c}_{i}:=c_i/2  \, , \quad \bar{c}_{12}:=c_{12}+\frac{1}{4}(c_1-c_2)^2 \, , \quad 
\bar{c}_{13}:=c_{13}+\frac{1}{4}(c_1-c_3)^2 \, , \quad 
\bar{c}_{23}:=c_{23}+\frac{1}{4}(c_2-c_3)^2 \, , \quad
\label{redef}
\end{equation}
\noindent we get the quadratic algebra:
\begin{equation}
f_{123}:=\frac{1}{2}\{\bar{c}_{12}, \bar{c}_{23}\}=\frac{1}{2}\{ \bar{c}_{23}, \bar{c}_{13}\}=\frac{1}{2}\{ \bar{c}_{13}, \bar{c}_{12}\} 
\label{f123}
\end{equation}
\begin{equation} \label{rel1}
\begin{split}
\{\bar{c}_{12}, f_{123}\}&=(\bar{c}_{23}-\bar{c}_{13})\bar{c}_{12}+(\bar{c}_1-\bar{c}_2)\bigl(c^{[3]}+(\bar{c}_1+\bar{c}_2) c^{[2]}-(\bar{c}_1+\bar{c}_2)^3\bigl) \\
\{\bar{c}_{13}, f_{123}\}&=(\bar{c}_{12}-\bar{c}_{23})\bar{c}_{13}+(\bar{c}_3-\bar{c}_1)\bigl(c^{[3]}+(\bar{c}_1+\bar{c}_3) c^{[2]}-(\bar{c}_1+\bar{c}_3)^3\bigl)\\
\{\bar{c}_{23}, f_{123}\}&=(\bar{c}_{13}-\bar{c}_{12})\bar{c}_{23}+(\bar{c}_2-\bar{c}_3)\bigl(c^{[3]}+(\bar{c}_2+\bar{c}_3) c^{[2]}-(\bar{c}_2+\bar{c}_3)^3\bigl)\\
\{\bar{c}_{12}, g_{123}\}&=2(\bar{c}_1-\bar{c}_2)f_{123}\\
\{\bar{c}_{13}, g_{123}\}&=2(\bar{c}_3-\bar{c}_1)f_{123}\\
\{\bar{c}_{23}, g_{123}\}&=2(\bar{c}_2-\bar{c}_3)f_{123}\\ 
\{f_{123},g_{123}\}&=(\bar{c}_1-\bar{c}_3)\bar{c}_{12}\bar{c}_{23}+(\bar{c}_3-\bar{c}_2)\bar{c}_{12}\bar{c}_{13}+(\bar{c}_2-\bar{c}_1)\bar{c}_{13}\bar{c}_{23}-(\bar{c}_1-\bar{c}_2)\times \\
&\quad (\bar{c}_2-\bar{c}_3)(\bar{c}_3-\bar{c}_1)\left(c^{[2]}-\frac{1}{3}\bigl((\bar{c}_1-\bar{c}_2)^2+(\bar{c}_1-\bar{c}_2)(\bar{c}_2-\bar{c}_3)+(\bar{c}_2-\bar{c}_3)^2\bigl)\right) \, .
\end{split}
\end{equation}

A quick check shows that $\{\bar{c}_{12}+\bar{c}_{13}+\bar{c}_{23}, P\}=0$ for $P=f_{123}$ and $P= g_{123}$, as expected. In this rescaled basis, the previous relations can be cast into the form:
\begin{equation} 
\begin{split}\label{eq1}
\bar{c}_1+\bar{c}_2+\bar{c}_3&=0,\\
\bar{c}_{12}+\bar{c}_{13}+\bar{c}_{23}&=c^{[2]}+\bar{c}_1^2+\bar{c}_2^2+\bar{c}_3^2,\\
g_{123}-\bar{c}_3 \bar{c}_{12}-\bar{c}_2 \bar{c}_{13}-\bar{c}_1 \bar{c}_{23}&=\frac{1}{2}c^{[3]}+\frac{5}{3}(\bar{c}_1^3+\bar{c}_2^3+\bar{c}_3^3) 
\end{split}
\end{equation}
together with:
\begin{equation}
g^2_{123}-f^2_{123}-\bigl(\bar{c}_{12}-(\bar{c}_1-\bar{c}_2)^2\bigl)\bigl(\bar{c}_{13}-(\bar{c}_1-\bar{c}_3)^2\bigl)\bigl(\bar{c}_{23}-(\bar{c}_2-\bar{c}_3)^2\bigl)=0 \, .\label{equ1}
\end{equation}
At this point, two comments are in order. First, we notice that from the functional relations it is possible to extrapolate the Casimir associated to the above quadratic algebra. In particular, from \eqref{eq1} we can obtain $g_{123}$ expressed in terms of the other generators and central elements. Then,  we can replace its expression into \eqref{equ1}. The resulting equation can be separated into two parts, with the Casimir written in terms of generators and central elements on the left-hand side, whereas on the right-hand side we obtain an expression depending only on the central elements. Explicitly, on the left-hand side we have 
\begin{equation}
\begin{split}
K &= f_{123}^2 + \bar{c}_{12} \bar{c}_{13} \bar{c}_{23}-\sum_{i\neq j\neq k} \bar{c}_k^2 \bar{c}_{ij}^2-\sum_{i\neq j\neq k} (  \bar{c}_i^2 +\bar{c}_j^2) \bar{c}_{ki} \bar{c}_{kj}\\
 & +\sum_{i\neq j\neq k}\left((\bar{c}_i - \bar{c}_j)^2 (\bar{c}_i - \bar{c}_k)^2 - \bar{c}_i\biggl( c^{[3]}+\frac{10}{3}(\bar{c}_1^3 + \bar{c}_2^3+ \bar{c}_3^3) \biggl)\right ) \bar{c}_{jk}, 
\end{split}
\end{equation}
while the right-hand side of the equation reads:
\begin{equation}
K=\frac{1}{4}\bigl(c^{[3]}\bigl)^2 + \frac{5}{3} (\bar{c}_1^3 +\bar{c}_2^3 +\bar{c}_3^3 )c^{[3]} +2(\bar{c}_1^2 +\bar{c}_2^2)(\bar{c}_1^2 +\bar{c}_3^2)(\bar{c}_2^2 +\bar{c}_3^2)     \, .
\label{cask}
\end{equation}
\noindent We further observe that, taking into account the functional relation \eqref{equ1}, we can restrict to consider just a subset of the relations \eqref{f123}-\eqref{rel1}. These share some similarities with the defining relations of the Racah algebra $R(3)$ \cite{bie2020racah, lat21}. In this setting, however, we see that the second-order and third-order Casimir of the Lie algebra appear on the right hand side of the relations. To the best of our knowledge, this represents a new Racah-type quadratic algebra, here arising from suitable combinations of the elements of the commutant of the Cartan subalgebra. We prove in the following that the well-known relations of Racah algebra $R(3)$ can be recovered after considering explicit canonical realizations for the generators, the latter being associated to the generic superintegrable system on the two-sphere after reduction.

\subsection{Connection with the superintegrable system on the two-sphere $\mathbb{S}^2$}

\noindent In order to provide an actual physical application to superintegrable systems we provide here a connection with the generic superintegrable system on the two-sphere  $\mathbb{S}^2$, determined by the Hamiltonian model (see \cite{Kalnins_2007}):
\begin{equation}
H=\frac{p_1^2}{2}+\frac{p_2^2}{2}+\frac{p_3^2}{2}+\frac{\alpha_1^2}{2s_1^2}+\frac{\alpha_2^2}{2s_2^2}+\frac{\alpha_3^2}{2s_3^2} = \frac{1}{2}\sum_{1 \leq i<j}^3(s_i p_j-s_j p_i)^2+\frac{1}{2}\sum_{i=1}^3\frac{\alpha_i^2}{s_i^2}\quad \cup \quad \begin{cases} s_1 p_1+s_2 p_2+s_3 p_3=0\\
s_1^2+s_2^2+s_3^2=1\end{cases} 
\label{super}
\end{equation}

 \noindent This problem has been faced in the paper \cite{Correa2020}, in relation to the Lie algebra $\mathfrak{su}(3)$ (see e.g. \cite{Calzada_2006}). In this context, the canonical realization for the generators of the underlying Lie algebra $\mathfrak{sl}(3)$ that allows us to connect with the generic superintegrable system on the sphere \eqref{super} reads:
\begin{equation}
h_1 ={\rm i}(\alpha_1-\alpha_2) \, , \quad h_2={\rm i}(\alpha_2-\alpha_3) \, , \quad
\begin{cases}
e_{ij}=-\frac{1}{2}\left((s_i p_j-s_j p_i)-{\rm i} \left(\alpha_i \frac{s_j}{s_i}+\alpha_j \frac{s_i}{s_j}\right) \right)  \\
e_{ji}=-\frac{1}{2}\left((s_j p_i-s_i p_j)-{\rm i} \left(\alpha_i \frac{s_j}{s_i}+\alpha_j \frac{s_i}{s_j}\right) \right) 
\end{cases}
\quad (1\leq i < j \leq 3) \, .
\label{eq:real}
\end{equation}
\noindent As a result of \eqref{eq:real}, we get the following identifications for the second and third-order Casimir invariants:
\begin{align}
c^{[2]}&=-\frac{1}{2}\left(H+\frac{1}{6}(\alpha_1+\alpha_2+\alpha_3)^2\right)\\
c^{[3]}&=\frac{{\rm i}}{3} (\alpha_1+\alpha_2+\alpha_3) c^{[2]}+\frac{{\rm i}}{27}(\alpha_1+\alpha_2+\alpha_3)^3  \, .
\end{align}
Notice that the third-order Casimir collapses to a combination of the second-order Casimir and the constants $\alpha_i$ appearing in the Hamiltonian, the latter being:
\begin{equation}
H=-2 c^{[2]}-\frac{1}{6}(\alpha_1+\alpha_2+\alpha_3)^2 \, .
\label{HS2}
\end{equation}
In terms of these elements, exactly the same relations given in \eqref{frel} are obtained. In particular, the element $g_{123}$ collapses as a consequence of the reduction:
\begin{equation}
g_{123}={\rm i}\; (\alpha_3 c_{12}+\alpha_2 c_{13}+\alpha_1 c_{23}+\alpha_1 \alpha_2 \alpha_3) \, .
\label{eq:g123red}
\end{equation}
This leads to the usual Racah algebra $R(3)$. In fact, we see that the elements $c_{ij}$ of the commutant of the Cartan subalgebra adopt the well-known expression:
\begin{equation}
c_{ij}=-\frac{1}{4}\left( (s_i p_j-s_j p_i)^2+\alpha_i^2 \frac{s_j^2}{s_i^2}+\alpha_j^2 \frac{s_i^2}{s_j^2}+2 \alpha_i \alpha_j\right) \,\quad 1\leq i < j \leq 3 
\label{commut}
\end{equation}
namely, they collapse to the constants of the motion associated to the model \eqref{HS2}. At this point, by performing the change of basis as in \eqref{redef}, re-expressing all in terms of the Hamiltonian $H$ and the rescaled constants of motion, that now read:
\begin{equation}
\bar{c}_{ij}=-\frac{1}{4}\left( (s_i p_j-s_j p_i)^2+(s_i^2+s_j^2)\left(\frac{\alpha_i^2}{s_i^2} +\frac{\alpha_j^2}{s_j^2} \right)\right) \,  \quad (1\leq i < j \leq 3)
\label{commutbij}
\end{equation}
\noindent we obtain the symmetry algebra associated to the generic superintegrable system on the sphere $\mathbb{S}^2$:

\begin{equation}\label{f123real}
\begin{split}
f_{123}&=\frac{1}{2}\{\bar{c}_{12}, \bar{c}_{23}\}=\frac{1}{2}\{ \bar{c}_{23}, \bar{c}_{13}\}=\frac{1}{2}\{ \bar{c}_{13}, \bar{c}_{12}\},\\
\{\bar{c}_{12}, f_{123}\}&=\bar{c}_{12}(\bar{c}_{23}-\bar{c}_{13})+\frac{1}{16}(\alpha_1^2-\alpha_2^2)(2 H-\alpha_3^2)\\
\{\bar{c}_{13},f_{123}\}&=\bar{c}_{13}(\bar{c}_{12}-\bar{c}_{23})+\frac{1}{16}(\alpha_3^2-\alpha_1^2)(2 H-\alpha_2^2)\\
\{\bar{c}_{23}, f_{123}\}&=\bar{c}_{23}(\bar{c}_{13}-\bar{c}_{12})+\frac{1}{16}(\alpha_2^2-\alpha_3^2)(2 H-\alpha_1^2) \, .
\end{split}
\end{equation}
along with the functional (actually linear) relation among the four constants of the motion:
\begin{equation}
\frac{H}{2}+\bar{c}_{12}+\bar{c}_{13}+\bar{c}_{23}+\frac{1}{4}(\alpha_1^2+\alpha_2^2+\alpha_3^2)=0 \, .
\label{eq:lineq}
\end{equation} 

\noindent Notice that the Casimir \eqref{cask} collapses to the following quadratic function of the Hamiltonian:

\begin{equation}
K=\Omega_1(\alpha_1, \alpha_2, \alpha_3) H^2+\Omega_2(\alpha_1, \alpha_2, \alpha_3) H+ \Omega_3(\alpha_1, \alpha_2, \alpha_3) 	\, ,
\label{EQ}
\end{equation}
where:
\begin{align}
\Omega_1(\alpha_1, \alpha_2, \alpha_3)=&-\frac{1}{144}(\alpha_1+\alpha_2+\alpha_3)^2\\
\Omega_2(\alpha_1, \alpha_2, \alpha_3)=&\frac{1}{144} \left(\alpha _1+\alpha _2+\alpha _3\right) \left((\alpha_1+\alpha_2+\alpha_3)^3-5\sum_{i\neq j\neq k} \alpha_i(\alpha_j+\alpha_k)^2+30 \alpha_1 \alpha_2 \alpha_3)\right)\\
\Omega_3(\alpha_1, \alpha_2, \alpha_3)=&\frac{1}{576} \bigl(\alpha _1^6-4 \left(\alpha _2+\alpha _3\right) \alpha _1^5+\left(9 \alpha _2^2+2 \alpha _3 \alpha _2+9 \alpha _3^2\right) \alpha _1^4-2 \left(\alpha _2+\alpha _3\right)\alpha_1^3\times \nonumber\\
& \left(4 \alpha _2^2-3 \alpha _3 \alpha _2+4 \alpha _3^2\right)+\left(9 \alpha _2^4-2 \alpha _3 \alpha _2^3+6 \alpha _3^2 \alpha _2^2-2 \alpha _3^3 \alpha _2+9 \alpha _3^4\right) \alpha _1^2 \nonumber \\
&  -2 \left(\alpha _2+\alpha _3\right) \left(\alpha _2^2+\alpha _3^2\right) \left(2 \alpha _2^2-3 \alpha _3 \alpha _2+2 \alpha _3^2\right) \alpha _1+\left(\alpha _2^2+\alpha _3^2\right)\times \nonumber \\
&   \left(\alpha _2^4-4 \alpha _3 \alpha _2^3+8 \alpha _3^2 \alpha _2^2-4 \alpha _3^3 \alpha _2+\alpha _3^4\right)\bigl) \, .
\end{align}
\noindent It is not surprising that in terms of the following quantities:
\begin{equation}
 C_{i}:=-\alpha_i^2/4 \, , \quad C_{ij}:=\bar{c}_{ij}  \, , \quad C_{123}:=-H/2 \, , \quad F_{123}:=f_{123}
\label{Racah}
\end{equation}
the relations of the rank-one Racah algebra $R(3)$ are satisfied \cite{bie2018racah, bie2020racah}:
\begin{equation}\label{eq:f123}
\begin{split}
F_{123}=& \frac{1}{2}\{C_{12},C_{23}\}=\frac{1}{2}\{C_{23},C_{13}\}=\frac{1}{2}\{C_{13},C_{12}\},\\
\{C_{12},F_{123}\}&=(C_{23}-C_{13})C_{12}+(C_2-C_1)(C_3-C_{123})\\
\{C_{13},F_{123}\}&=(C_{12}-C_{23})C_{13}+(C_1-C_3)(C_2-C_{123})\\
\{C_{23},F_{123}\}&=(C_{13}-C_{12})C_{23}+(C_3-C_2)(C_1-C_{123}) \, ,
\end{split}
\end{equation}
together with the linear relation
\begin{equation}
C_{123}=C_{12}+C_{13}+C_{23}-C_1-C_2-C_3. 
\label{RacahLinRel}
\end{equation}
The latter is obtained after considering the explicit realisation in terms of canonical coordinates $(s_i,p_i)$. An alternative presentation, that will turn out to be the most suitable  for higher-rank cases, is the one involving the generators \cite{centr}:
\begin{equation}
P_{ii}:=2 C_i  \,  \qquad P_{ij}:=C_{ij}-C_i-C_j \, ,
\label{central}
\end{equation}
explicitly, in the given canonical realisation:
\begin{equation}
P_{ii}=-\alpha_i^2/2  \, ,\qquad P_{ij}=-\frac{1}{4}\left( (s_i p_j-s_j p_i)^2+\left(\alpha_i^2\frac{s_j^2}{s_i^2} +\alpha_j^2\frac{s_i^2}{s_j^2} \right)\right) \, .
\label{centr}
\end{equation}
\noindent In terms of these generators, the rank-one Racah algebra $R(3)$ can be rewritten as:
\begin{equation}\label{eq:f123P}
\begin{split}
F_{123}=& \frac{1}{2}\{P_{12},P_{23}\}=\frac{1}{2}\{P_{23},P_{13}\}=\frac{1}{2}\{P_{13},P_{12}\},\\
\{P_{12},F_{123}\}&=(P_{12}+P_{11})P_{23}-(P_{12}+P_{22})P_{13}\\
\{P_{13},F_{123}\}&=(P_{13}+P_{33})P_{12}-(P_{13}+P_{11})P_{23}\\
\{P_{23},F_{123}\}&=(P_{23}+P_{22})P_{13}-(P_{23}+P_{33})P_{12} \, ,
\end{split}
\end{equation}
where the Hamiltonian of the model is given by 
\begin{equation}
H=-2\sum_{1 \leq i<j}^3P_{ij}-\sum_{i=1}^3P_{ii}\, .
\label{Ham}
\end{equation}

\noindent At this stage, it is worthy to be mentioned that the algebra can be rewritten in a more compact form if one takes into account the symmetry properties of the generators with respect to the indices. In fact, considering that $P_{ij}=P_{ji}$ is symmetric on exchange of two indices, whereas the three-indices generator $F_{ijk}=-F_{jik}=F_{jki}$ is antisymmetric, for $i \neq j \neq k \in \{1,2,3\}$, the algebra can be rewritten as \cite{centr}:
\begin{equation}
\{P_{ij}, P_{jk}\}=2 F_{ijk}\, , \qquad \{P_{jk}, F_{ijk}\}=(P_{jk}+P_{jj})P_{ik}-(P_{jk}+P_{kk})P_{ij} \, .
\label{eq:ants}
\end{equation}

\subsubsection{The general form of the quadratic algebra}

\noindent In analogy with the Racah case, we can use the symmetry properties on the indices of generators in $C_{\mathfrak{sl}(3)^{\ast}}(\mathfrak{h}^{\ast})$
to present the associated quadratic algebra in a more general form.

\bigskip The defining elements of the quadratic algebra are defined for $i \neq j \neq k \in \{1,2,3\}$ as:
\begin{equation}
c_i:=\frac{1}{3}\sum_{j=1}^2 (3-j)h_j-\sum_{j=1}^{i-1} h_j \qquad c_{ij}:=p_{i,j}, \qquad f_{ijk}:=\frac{1}{2}(p_{i,k,j}-p_{i,j,k})  \qquad g_{ijk}:=\frac{1}{2}(p_{i,k,j}+p_{i,j,k}) \, .
\end{equation}

\noindent The elements $c_i$ play the role of central elements of the algebra. In particular, they depend on the Cartan elements $\{h_1, h_2\}$ in such a way $c_1+c_2+c_3=0$. The two and three indices elements satisfy the following symmetry properties:
\begin{align}
&c_{ij}=c_{ji} \hskip 3cm (\text{symmetric}) \\
&f_{ijk}=-f_{jik}=f_{jki} \hskip 1.35cm  (\text{skew-symmetric})\\
&g_{ijk}=g_{jik}=g_{jki} \hskip 1.65cm  (\text{symmetric})
\end{align}
\medskip
\noindent The previous elements, for $i \neq j \neq k \in \{1,2,3\}$ satisfy the following quadratic algebra:
\begin{equation}\label{equafin} 
\begin{split}
\{c_i, \cdot\}&=0\\
\{c_{ij},c_{jk}\}&=2 f_{ijk}\\
\{c_{jk},f_{ijk}\}&=(c_{ik}-c_{ij})c_{jk}+(c_j-c_k)g_{ijk}\\
\{c_{jk},g_{ijk}\}&=(c_j-c_k)f_{ijk}\\
\{f_{ijk},g_{ijk}\}&=\frac{1}{2}\bigl((c_i-c_k)c_{ij}c_{jk}+(c_k-c_j)c_{ki}c_{ij}+(c_j-c_i)c_{jk}c_{ki} \bigl) 
\end{split}
\end{equation}
together with the additional functional relation:
\begin{equation}\label{addrel}
f_{ijk} f_{kji}+g_{ijk} g_{kji}=c_{ij} c_{jk} c_{ki} \, .
\end{equation}
With this notation, the Casimir invariant of the quadratic algebra can be expressed as:
\begin{equation}
K_{ijk}=f_{ijk}f_{jki} +c_{ij}c_{jk}c_{ki}-(c_i c_{jk}+c_j c_{ik}+c_k c_{ij})\left(g_{ijk}-\frac{1}{4}(c_i c_{jk}+c_j c_{ik}+c_k c_{ij})\right)\, ,
\label{casim}
\end{equation}
where the symmetry properties $K_{ijk}=K_{jik}=K_{jki}$ hold. In particular, 
\begin{equation}
K_{ijk}=\frac{1}{4}\bigl(c^{[3]}-c_i c_j c_k\bigl)^2 \, .
\label{cascs}
\end{equation}

\subsection{Systems on the 3-sphere and $\mathfrak{sl}(4)$}
\noindent We next consider the rank three Lie algebra  $\mathfrak{sl}(4)$. 
 In this case, the system (\ref{CAR}) has $12$ functionally independent solutions, while the centralizer $C_{S(\mathfrak{sl}(4))}(\mathfrak{h})$ is of dimension $23$. Following (\ref{dice}), a basis of $C_{\mathfrak{sl}(4)^{\ast}}(\mathfrak{h}^{\ast})$ is given by 
\begin{equation}\label{gen4}
\begin{split}
h_1,\quad h_2,\quad h_3,\quad p_{1,2},\quad p_{1,3},\quad p_{1,4},\quad p_{2,3},\quad p_{2,4},\quad p_{3,4},\quad p_{1,2,3},\quad p_{1,2,4},\quad p_{1,3,4},\quad p_{2,3,4}, \\
 p_{1,3,2},\quad p_{1,4,2},\quad p_{1,4,3},\quad p_{2,4,3},\quad p_{1,2,3,4},\quad p_{1,2,4,3},\quad p_{1,3,2,4},\quad 
p_{1,4,2,3},\quad p_{1,4,3,2},\quad p_{1,3,4,2},\\
\end{split}
\end{equation}
with $\left\{h_1, h_2, h_3, p_{1,2}, p_{1,3},p_{1,4}, p_{2,3}, p_{2,4}, p_{3,4},p_{1,2,3}, p_{1,2,4},p_{1,3,4}\right\}$ being functionally independent. 

\medskip
In analogy to the previous case, we can use symmetry properties on the indices of the generators to determine a more suitable basis that simplifies the expression of the 
Lie-Poisson brackets and results in a more transparent description of the polynomial algebra generated by the monomials in (\ref{gen4}). We redefine the basis 
$i \neq j \neq k \neq l \in \{1,2,3,4\}$ as follows:
 \begin{equation}
 c_i:=\frac{1}{4}\sum_{j=1}^3 (4-j)h_j-\sum_{j=1}^{i-1} h_j, \quad c_{ij}:=p_{i,j}, \quad f_{ijk}:=\frac{1}{2}(p_{i,k,j}-p_{i,j,k}),  \quad g_{ijk}:=\frac{1}{2}(p_{i,k,j}+p_{i,j,k})
 \end{equation}
Again, the one index elements $c_i$ play the role of central elements for the algebra, and fulfill the linear relation $c_1+c_2+c_3+c_4=0$. The $c_{ij}$ and $g_{ijk}$ are symmetric, while the $f_{ijk}$ are skew-symmetric. For elements depending on four indices, instead of using the symmetric and skew-symmetric counterparts, it is computationally convenient to consider the elements 
\begin{equation}
f_{ijkl}:=\frac{1}{2}(p_{i,l,k,j}-p_{i,j,k,l}),  \qquad g_{ijkl}:=\frac{1}{2}(p_{i,l,k,j}+p_{i,j,k,l}). 
\end{equation}
These four-indices elements $f_{ijkl}$ and $g_{ijkl}$ satisfy the following symmetry properties:
 \begin{equation}\label{sec}
 f_{ijkl}=-f_{jilk},\quad  g_{ijkl}=g_{jilk},
 \end{equation}
meaning that they are antisymmetric and symmetric with respect to the simultaneous exchange of the first and second pair of indices, and of the central pair with the external one, respectively. Besides, they both satisfy the cyclic property. The composition of these two properties provides several equalities for the four indices elements. Those equalities will be taken into account when presenting the algebra in general form.

\medskip
\noindent For $i \neq j \neq k \neq l \in \{1,2,3,4\}$, in terms of the previously defined elements, the following cubic algebra relations are obtained. The elements with one index are central, hence $\{c_i,\cdot\}=0$. For the case of two indices elements we get
\begin{equation} 
\begin{split}\label{cub1}
\{c_{ij},c_{kl}\}&=0,\quad  \{c_{ij},c_{jk}\}=2 f_{ijk}, 
\end{split}
\end{equation}
whereas for the bracket of quadratic and cubic elements we obtain 
\begin{equation} 
\begin{split}
  \{c_{jk},f_{ijk}\}&=(c_{ik}-c_{ij})c_{jk}+(c_j-c_k)g_{ijk},\\
 \{c_{jk},g_{ijk}\}&=(c_j-c_k)f_{ijk},\\
 \{c_{kl},f_{ijk}\}&=g_{i j l k}-g_{i j k l},\\
 \{c_{kl},g_{ijk}\}&=f_{i j l k}-f_{i j k l},\\
 \{c_{jk},f_{ijk}\}&=(c_{ik}-c_{ij})c_{jk}+(c_j-c_k)g_{ijk}\\
 \{c_{jk},g_{ijk}\}&=(c_j-c_k)f_{ijk}\\
 \{c_{kl},f_{ijk}\}&=g_{i j l k}-g_{i j k l}\\
 \{c_{kl},g_{ijk}\}&=f_{i j l k}-f_{i j k l}.   
\end{split}
\end{equation}
We next consider the brackets of elements with three indices:
\begin{equation} 
\begin{split}
\{f_{ijk},g_{ijk}\}&=\frac{1}{2}\bigl((c_i-c_k)c_{ij}c_{jk}+(c_k-c_j)c_{ki}c_{ij}+(c_j-c_i)c_{jk}c_{ki} \bigl),\\
 \{f_{ijk},f_{jkl}\}&=\frac{1}{2}((c_{ij}-c_{ki})f_{jkl}+(c_{kl}-c_{jl})f_{ijk}+(f_{ilj}+f_{ilk})c_{jk}+(c_j-c_k)f_{ijlk}),\\
 \{g_{ijk},g_{jkl}\}&=\frac{1}{2}((c_{ij}-c_{ki})f_{jkl}+(c_{kl}-c_{jl})f_{ijk}+(f_{ijl}+f_{ikl})c_{jk}+(c_j-c_k)f_{iklj}),\\
 \{f_{ijk},g_{jkl}\}&=\frac{1}{2}(c_{ij}-c_{ki})g_{jkl}+(c_{kl}-c_{jl})g_{ijk}+(g_{ijl}-g_{ikl})c_{jk}+(c_k-c_j)g_{iklj}), . 
\end{split}
\end{equation}
Elements with two indices with those having four indices close as
\begin{equation} 
\begin{split}
 \{c_{kl},f_{ijkl}\}&=(g_{ilj}-g_{ikj})c_{kl}+(c_k-c_l)g_{ijkl},\\
 \{c_{kl},f_{ijlk}\}&=(g_{ikj}-g_{ilj})c_{kl}+(c_l-c_k)g_{ijlk},\\
 \{c_{kl},f_{iljk}\}&=(c_{jl}-c_{jk})g_{ikl} + (c_{ik}-c_{il})g_{jkl}, \\
 \{c_{kl},g_{ijkl}\}&=(f_{ijl}-f_{ijk})c_{kl}+(c_k-c_l)f_{ijkl},\\
 \{c_{kl},g_{ijlk}\}&=(f_{ijk}-f_{ijl})c_{kl}+(c_l-c_k)f_{ijlk},\\
 \{c_{kl},g_{iljk}\}&=(c_{jk}-c_{jl})f_{ikl} +(c_{ik}-c_{il})f_{jkl}, 
\end{split}
\end{equation}
while for elements with three and four indices, respectively, we get the relations 
\begin{equation*} 
\begin{split}
\{f_{jkl},f_{ijkl}\}=&\frac{1}{2}\bigl((c_{jk}-c_{kl})f_{ijkl}+(g_{ikl}-g_{ijk})f_{jkl}+(f_{ikl}-f_{ijk})g_{jkl}+\bigl((c_l-c_k)c_{jk}+(c_k-c_j)c_{kl}\bigl)f_{ijl}\bigl),\\
 \{f_{jkl},f_{ijlk}\}= &\frac{1}{2}\bigl((c_{kl} - c_{jl}) f_{ijlk} + (g_{ikl} - g_{ijl})  f_{jkl}+ (f_{ijl} + 
 f_{ikl}) g_{jkl} + ( (c_j - c_l) c_{kl} + (c_l - c_k) c_{jl}) f_{ijk}\bigl),\\ 
\{f_{jkl},f_{iljk}\}=&\frac{1}{2}\bigl((c_{jl} - c_{jk}) f_{iljk} +  (g_{ijk} - g_{ijl}) f_{jkl}+ (f_{ijk} + f_{ijl}) g_{jkl} + ( (c_j - c_k) c_{jl} + (c_l - c_j) c_{jk}) f_{ikl}),\\
\{g_{jkl},g_{ijkl}\}=&\frac{1}{2}\bigl((c_{jk}-c_{kl})f_{ijkl}+(g_{ikl}-g_{ijk})f_{jkl}+(f_{ikl}-f_{ijk})g_{jkl}+\bigl((c_k-c_l)c_{jk}+(c_j-c_k)c_{kl}\bigl)f_{ijl}\bigl),\\
 \{g_{jkl},g_{ijlk}\}=&\frac{1}{2} \bigl((c_{jl} - c_{kl}) f_{ijlk} + (g_{ijl} - g_{ikl}) f_{jkl} - (f_{ijl} + f_{ikl}) g_{jkl} + ( (c_j - c_l) c_{kl} + (c_l - c_k) c_{jl}) f_{ijk}\bigl),
\end{split}
\end{equation*}

\begin{equation*} 
\begin{split}
\{g_{jkl},g_{iljk}\}=&\frac{1}{2} \bigl((c_{jl} - c_{jk}) f_{iljk} +  (g_{ijk} - g_{ijl})f_{jkl}+ (f_{ijk} + f_{ijl}) g_{jkl} + ( (c_k - c_j) c_{jl} + (c_j - c_l) c_{jk}) f_{ikl}),\\
\{f_{jkl},g_{ijkl}\}=&\frac{1}{2}\left((c_{jk} - c_{kl}) g_{ijkl} +  (c_{ij}-c_{il})c_{jk}c_{kl}+(f_{ikl} - f_{ijk})f_{jkl}+ (g_{ikl} - g_{ijk}) g_{jkl} + ( (c_k - c_j) c_{kl} \right.\\ 
 & +\left.(c_l - c_k) c_{jk}) g_{ijl}\right),\\
 \{f_{jkl},g_{ijkl}\}=&\frac{1}{2}\left((c_{jk} - c_{kl}) g_{ijkl} +  (c_{ij}-c_{il})c_{jk}c_{kl}+(f_{ikl} - f_{ijk})f_{jkl}+ (g_{ikl} - g_{ijk}) g_{jkl} + ( (c_k - c_j) c_{kl} \right.\\
 & \left.+ (c_l - c_k) c_{jk}) g_{ijl}\right),\\ 
\{f_{jkl},g_{ijlk}\}=&\frac{1}{2}\left((c_{kl} - c_{jl}) g_{ijlk} +  (c_{ik}-c_{ij})c_{jl}c_{kl}-(f_{ijl} + f_{ikl})f_{jkl}+ (g_{ijl} - g_{ikl}) g_{jkl} + ( (c_j - c_l) c_{kl} \right.\\
& \left.+ (c_l - c_k) c_{jl}) g_{ijk}\right),\\
\{f_{jkl},g_{iljk}\}=&\frac{1}{2}\left((c_{jl} - c_{jk}) g_{iljk} +  (c_{il}-c_{ik})c_{jk}c_{jl}+(f_{ijk} + f_{ijl})f_{jkl}+ (g_{ijk} - g_{ijl}) g_{jkl} + ( (c_j - c_l) c_{jk} \right.\\
& \left.+ (c_k - c_l) c_{jl}) g_{ikl}\right),\\
\{g_{jkl},f_{ijkl}\}=&\frac{1}{2}\left((c_{jk} - c_{kl}) g_{ijkl} +  (c_{il}-c_{ij})c_{jk}c_{kl}+(f_{ikl} - f_{ijk})f_{jkl}+ (g_{ikl} - g_{ijk}) g_{jkl} + ( (c_j - c_k) c_{kl} \right.\\
& \left. + (c_k - c_l) c_{jk}) g_{ijl}\right),\\
\{g_{jkl}, f_{ijlk}\}=&\frac{1}{2}\left((c_{jl} - c_{kl}) g_{ijlk} +  (c_{ik}-c_{ij})c_{jl}c_{kl}+(f_{ijl} + f_{ikl})f_{jkl}+ (g_{ikl} - g_{ijl}) g_{jkl} + ( (c_j - c_l) c_{kl}\right.\\
& \left.+ (c_l - c_k) c_{jl}) g_{ijk}\right),\\
\{g_{jkl},f_{iljk}\}=&\frac{1}{2}\left((c_{jl} - c_{jk}) g_{iljk} +  (c_{il}-c_{ik})c_{jk}c_{jl}+(f_{ijk} + f_{ijl})f_{jkl}+ (g_{ijk} - g_{ijl}) g_{jkl} + ( (c_j - c_l) c_{jk} \right.\\
& \left.+ (c_k - c_l) c_{jl}) g_{ikl}\right).
\end{split}
\end{equation*}
Finally, the brackets of elements with four indices elements lead to the following identities
\begin{equation*} 
{\small
\begin{split}
 \{f_{ijkl},f_{ijlk}\}=&\frac{1}{2}\left( \bigl((c_{jk} + c_{jl} - c_{kl}) f_{ikl} + ( c_{kl}-c_{ik}- c_{il}) f_{jkl}\bigl) c_{ij} + \bigl((c_{ij} - 
 c_{il}  - c_{jl} ) f_{ijk} + (c_{ik}  + c_{jk} - c_{ij}) f_{ijl}\bigl) c_{kl} \right. \\
 & \left. + (c_i - c_j) (f_{ikl} g_{jkl}+f_{jkl} g_{ikl} ) + (c_l - c_k) (f_{ijk} g_{ijl}+f_{ijl} g_{ijk} )\right),\\
 \{f_{ijkl},f_{iljk}\}=&\frac{1}{2}\left( \bigl((c_{ik} + c_{kl} - c_{il}) f_{ijl} + ( c_{il}-c_{ij}- c_{jl}) f_{ikl}\bigl) c_{jk} + \bigl((c_{jl} + 
 c_{kl}  - c_{jk} ) f_{ijk} + (c_{jk}  - c_{ij} - c_{ik}) f_{jkl}\bigl) c_{il} \right.\\
 & \left. + (c_i - c_l) (f_{ijk} g_{jkl}+f_{jkl} g_{ijk} ) + (c_j - c_k) (f_{ikl} g_{ijl}+f_{ijl} g_{ikl} )\right), \\
\{f_{ijlk},f_{iljk}\}=&\frac{1}{2} \left( \bigl((c_{ik} - c_{il} - c_{kl}) f_{ijk} + ( c_{ik}-c_{ij}- c_{jk}) f_{ikl}\bigl) c_{jl} + \bigl((c_{jl} -c_{jk}  - c_{kl} ) f_{ijl} + (c_{jl}  - c_{ij} - c_{il}) f_{jkl}\bigl) c_{ik} \right. \\
 & \left. + (c_i - c_k) (f_{jkl} g_{ijl}-f_{ijl} g_{jkl} ) + (c_l - c_j) (f_{ijk} g_{ikl}-f_{ikl} g_{ijk} )\right),\\
 \{g_{ijkl}, g_{ijlk}\}=&\frac{1}{2}\left( \bigl((c_{kl} - c_{jk} - c_{jl}) f_{ikl} + ( c_{ik} + c_{il}-c_{kl}) f_{jkl}\bigl) c_{ij} + \bigl((c_{ij} -c_{il} - c_{jl}) f_{ijk} + (c_{ik} + c_{jk} - c_{ij}) f_{ijl}\bigl) c_{kl}  \right.\\
  & \left. + (c_j - c_i) (f_{ikl} g_{jkl} + f_{jkl} g_{ikl}) + (c_l - c_k) ( f_{ijl} g_{ijk} + f_{ijk} g_{ijl})\right),\\ 
 \{g_{ijkl},g_{iljk}\}=&\frac{1}{2}\left( \bigl((c_{il} - c_{ik} - c_{kl}) f_{ijl} + ( c_{ij} + c_{jl}-c_{il}) f_{ikl}\bigl) c_{jk} + \bigl((c_{jl} +c_{kl} - c_{jk}) f_{ijk} + (c_{jk} - c_{ij} - c_{ik}) f_{jkl}\bigl) c_{il}\right.  \\
  & \left. + (c_i - c_l) (f_{ijk} g_{jkl} + f_{jkl} g_{ijk} ) + (c_k - c_j) ( f_{ikl} g_{ijl} + f_{ijl} g_{ikl})\right),\\  
\{g_{ijlk},g_{iljk}\}=&\left( \bigl((c_{ik} - c_{il} - c_{kl}) f_{ijk} + ( c_{ik} - c_{ij}-c_{jk}) f_{ikl}\bigl) c_{jl} + \bigl((c_{jk} +c_{kl} - c_{jl}) f_{ijl} + (c_{ij} + c_{il} - c_{jl}) f_{jkl}\bigl) c_{ik} \right. \\
  & \left. + (c_k - c_i) (f_{jkl} g_{ijl} - f_{ijl} g_{jkl} ) + (c_l - c_j) ( f_{ijk} g_{ikl} - f_{ikl} g_{ijk})\right),\\
\{f_{ijkl},g_{ijkl}\}=& \frac{1}{2}\bigl( \bigl((c_i - c_l) c_{ij} - (c_i - c_j) c_{il}\bigl)c_{jk} c_{kl} +\bigl((c_l - c_k) c_{jk} + (c_k - c_j) c_{kl}\bigl)c_{ij}c_{il} \bigl), \\
\{f_{ijkl},g_{ijlk}\}=&\frac{1}{2} \left( \bigl(( c_{kl}-c_{jk} - c_{jl}) g_{ikl} + ( c_{ik} + c_{il} - c_{kl}) g_{jkl}\bigl) c_{ij}+\bigl((c_{ij} - c_{il} - c_{jl}) g_{ijk} + ( c_{il} + c_{jk}-c_{ij} ) g_{ijl}\bigl) c_{kl}  \right. \\
 & \left. + (c_l - c_k) (f_{ijk} f_{ijl} + g_{ijk} g_{ijl}) + (c_j - c_i) (f_{ikl} f_{jkl} + g_{ikl} g_{jkl})\right),\\
\{f_{ijkl},g_{iljk}\}=&\frac{1}{2}\left( \bigl(( c_{il}-c_{ik} - c_{kl}) g_{ijl} + ( c_{ij} - c_{il} + c_{jl}) g_{ikl}\bigl) c_{jk}+\bigl((c_{jl} + c_{kl} - c_{jk}) g_{ijk} + ( c_{jk} - c_{ij}-c_{ik} ) g_{jkl}\bigl) c_{il}  \right. \\
  & \left. + (c_k - c_j) (f_{ijl} f_{ikl} + g_{ijl} g_{ikl}) + (c_i - c_l) (f_{ijk} f_{jkl} + g_{ijk} g_{jkl})\right),\\
\{f_{ijlk},g_{ijkl}\}=& \frac{1}{2}\left( \bigl((c_{kl}-c_{jk} - c_{jl}) g_{ikl} + (c_{ik} + c_{il} - c_{kl}) g_{jkl}\bigl) c_{ij} + \bigl((c_{il} +
   c_{jl} - c_{ij}) g_{ijk} + (c_{ij} - c_{ik} - c_{jk}) g_{ijl}\bigl) c_{kl} \right. \\
   &\left. + (c_k - c_l) (f_{ijk} f_{ijl} + g_{ijk} g_{ijl}) +(c_j - c_i) (f_{ikl} f_{jkl} + g_{ikl} g_{jkl})\right),\\
\{f_{ijlk},g_{iljk}\}=& \frac{1}{2} \left( \bigl((c_{ik} - c_{il} - c_{kl}) g_{ijk} + (c_{ij} - c_{ik} + c_{jk}) g_{ikl}\bigl) c_{jl} + \bigl((c_{jk} - c_{jl} + c_{kl}) g_{ijl} + ( c_{jl}-c_{ij} - c_{il}) g_{jkl}\bigl) c_{ik} \right. \\
  & \left. + (c_j - c_l) (f_{ijk} f_{ikl} - g_{ijk} g_{ikl}) + (c_k - c_i) (f_{ijl} f_{jkl} - g_{ijl} g_{jkl})\right),\\
\end{split} }
\end{equation*}

\begin{equation*} 
\begin{split}
\{f_{iljk},g_{ijkl}\}=&\frac{1}{2} \left( \bigl((c_{jk} - c_{jl} - c_{kl}) g_{ijk} + (c_{ij} + c_{ik} - c_{jk}) g_{jkl}\bigl) c_{il} + \bigl((c_{il} - c_{ik} - c_{kl}) g_{ijl} + (c_{ij} - c_{il} + c_{jl}) g_{ikl}\bigl) c_{jk} \right. \\
   &\left. + (c_l - c_i) (f_{ijk} f_{jkl} + g_{ijk} g_{jkl}) + (c_k - c_j) (f_{ijl} f_{ikl} + g_{ijl} g_{ikl})\right),\\
\{f_{iljk},g_{ijlk}\}=& \frac{1}{2}\left( \bigl((c_{jk} - c_{jl} + c_{kl}) g_{ijl} + (  
       c_{jl}-c_{ij} - c_{il}) g_{jkl}\bigl) c_{ik} + \bigl((c_{il} + c_{kl}-c_{ik}) g_{ijk} + (c_{ik}-c_{ij} - c_{jk}) g_{ikl}\bigl) c_{jl} \right. \\
 & \left. +(c_l - c_j) (f_{ijk} f_{ikl} - g_{ijk} g_{ikl}) + (c_k - c_i) ( f_{jkl} f_{ijl} - g_{jkl} g_{ijl})\right).
\end{split}
\end{equation*}
 
\medskip
\noindent  Besides the dependence relations enumerated in (\ref{rela1}), there are some additional functional identities like  
\begin{equation} \label{addrels}
\begin{split}
 f_{ijk} f_{kji}+g_{ijk} g_{kji}&=c_{ij} c_{jk} c_{ki}, \\
 f_{ijkl} f_{lkji}+g_{ijkl} g_{lkji}&=c_{ij} c_{jk} c_{kl} c_{li}.  
\end{split}
\end{equation}

\noindent Once we have described the general structure of the cubic algebra $\mathcal{A}_4$, we next show that, once a realization associated to the generic superintegrable systems on the three-sphere $\mathbb{S}^3$ is considered, the cubic structure reduces to that of the quadratic algebra $R(4)$.

\subsection{Connection with the superintegrable system on the three-sphere $\mathbb{S}^3$}

\noindent The superintegrable system on the three sphere  $\mathbb{S}^3$ is defined by the Hamiltonian:
\begin{equation}
H=\frac{p_1^2}{2}+\frac{p_2^2}{2}+\frac{p_3^2}{2}+\frac{p_4^2}{2}+\frac{\alpha_1^2}{2s_1^2}+\frac{\alpha_2^2}{2s_2^2}+\frac{\alpha_3^2}{2s_3^2}+\frac{\alpha_4^2}{2s_4^2} = \frac{1}{2}\sum_{1 \leq i<j}^4(s_i p_j-s_j p_i)^2+\frac{1}{2}\sum_{i=1}^4\frac{\alpha_i^2}{s_i^2}  \label{superr}
\end{equation}
subjected to the constraints 
\begin{equation}
s_1 p_1+s_2 p_2+s_3 p_3+s_4 p_4=0,\quad s_1^2+s_2^2+s_3^2+s_4^2=1.
\end{equation}

\noindent In analogy with the rank-two case, we consider the canonical realization for the generators of $\mathfrak{sl}(4)$:
\begin{equation}\label{eq:realsl4}
\begin{split}
h_1 ={\rm i}(\alpha_1-\alpha_2) \, , \quad h_2={\rm i}(\alpha_2-\alpha_3) \, , \quad  h_3={\rm i}(\alpha_3-\alpha_4), \\
 e_{ij}=-\frac{1}{2}\left((s_i p_j-s_j p_i)-{\rm i} \left(\alpha_i \frac{s_j}{s_i}+\alpha_j \frac{s_i}{s_j}\right) \right),  \\[0.05cm]
 e_{ji}=-\frac{1}{2}\left((s_j p_i-s_i p_j)-{\rm i} \left(\alpha_i \frac{s_j}{s_i}+\alpha_j \frac{s_i}{s_j}\right) \right).
\end{split}
\end{equation}
From this realization, it is straightforward to derive the following identifications with the quadratic, cubic and quartic Casimir invariants of $\mathfrak{sl}(4)^{\ast}$:
\begin{equation} 
\begin{split}
c^{[2]}=&-\frac{1}{2}\left(H+\frac{1}{4}(\alpha_1+\alpha_2+\alpha_3+\alpha_4)^2\right)\\
c^{[3]}=&\frac{{\rm i}}{2} (\alpha_1+\alpha_2+\alpha_3+\alpha_4) c^{[2]}+\frac{{\rm i}}{16}(\alpha_1+\alpha_2+\alpha_3+\alpha_4)^3  \\
c^{[4]}=&-\frac{1}{16} \left(\alpha _1+\alpha _2+\alpha _3+\alpha _4\right){}^2 c^{[2]}+\frac{1}{8} \left(\alpha _1+\alpha _2+\alpha _3\right) \left(\alpha _1+\alpha _2+\alpha _3+\alpha _4\right){}^3\\
& -\frac{1}{8} \left((\alpha _1+\alpha _2+\alpha _3)^2+3 \left(\alpha _1 \alpha _2+\alpha _2 \alpha _3+\alpha _1 \alpha _3\right)\right) \left(\alpha _1+\alpha _2+\alpha _3+\alpha _4\right){}^2\\
& +\frac{1}{2} (\alpha_1+\alpha_2)(\alpha_2+\alpha_3)(\alpha_1+\alpha_3) \left(\alpha _1+\alpha _2+\alpha _3+\alpha _4\right)-\frac{9}{256}  \left(\alpha _1+\alpha _2+\alpha _3+\alpha _4\right){}^4\\
& +2 \alpha _1 \alpha _2 \alpha _3 \alpha_4.
\end{split}
\end{equation}
We observe that both the third-order and fourth-order Casimir invariants collapse to a combination of the quadratic one and the constants $\alpha_i$ appearing in the Hamiltonian, the latter being given by:
\begin{equation}
H=-2 c^{[2]}-\frac{1}{4}(\alpha_1+\alpha_2+\alpha_3+\alpha_4)^2 \, .
\end{equation}
Again, the two indices elements $c_{ij}$ belonging to the commutant of the Cartan subalgebra assume the well-known expression:
\begin{equation}
c_{ij}=-\frac{1}{4}\left( (s_i p_j-s_j p_i)^2+\alpha_i^2 \frac{s_j^2}{s_i^2}+\alpha_j^2 \frac{s_i^2}{s_j^2}+2 \alpha_i \alpha_j\right) \, \qquad (1 \leq i<j \leq 4) \, .
\label{commuta}
\end{equation}
 These elements, together with the elements $\{f_{ijk}, g_{ijk}\}$ and $\{f_{ijkl}, g_{ijkl}\}$, reproduce exactly the relations of the cubic algebra $\mathcal{A}_4$ presented in the previous paragraph (equations \eqref{cub1} onwards) and the related functional relations. The crucial point in this construction is that the $g_{ijk}$, as well as the four-indices elements $f_{ijkl},g_{ijkl}$ are all expressible in terms of elements with a lower number of indices after the reduction to canonical coordinates. More explicitly, the following identities are satisfied for any $i \neq j \neq k \neq l \in \{1,2,3,4\}$:
 \begin{equation} \label{collapses}
\begin{split}
g_{ijk}=&\frac{{\rm i}}{2}(\alpha_k c_{ij}+\alpha_j c_{ik}+\alpha_i c_{jk}+\alpha_i \alpha_j \alpha_k), \\
g_{ijkl}=&\frac{1}{2}(c_{ij}c_{kl}+c_{il}c_{jk}-c_{ik}c_{jl}-\alpha_i \alpha_k c_{jl}-\alpha_j \alpha_l c_{ik}-\alpha_i \alpha_j \alpha_k \alpha_l), \\
f_{ijkl}=&\frac{{\rm i}}{2}(\alpha_l f_{ijk}+\alpha_k f_{ijl}+\alpha_j f_{ikl}+\alpha_i f_{jkl}).
\end{split}
\end{equation}
This means specifically that, as a result of the canonical realization, the cubic algebra collapses to a quadratic one with basis elements:
 \begin{equation}
 \alpha_i,\; c_{ij},\; f_{ijk},\; H.
 \end{equation} 
However remarkable this result may appear, it is albeit expected, as the algebra associated to the generic superintegrable system on the three-sphere $\mathbb{S}^3$, the rank-two Racah algebra $R(4)$, is quadratic. To make this connection with $R(4)$ more explicit, we introduce the constants of the motion
 \begin{equation}
   \bar{c}_{ij}:=-\frac{1}{4}\left( (s_i p_j-s_j p_i)^2+(s_i^2+s_j^2)\left(\frac{\alpha_i^2}{s_i^2} +\frac{\alpha_j^2}{s_j^2} \right)\right),  \qquad (1\leq i < j \leq 4) \, ,
 \label{redefg}
 \end{equation}
such that the following constraint holds:
\begin{equation}
\frac{H}{2}+\bar{c}_{12}+\bar{c}_{13}+\bar{c}_{14}+\bar{c}_{23}+\bar{c}_{24}+\bar{c}_{34}+\frac{1}{2}(\alpha_1^2+\alpha_2^2+\alpha_3^2+\alpha_4^2)=0 \, .
\label{eq:linequa}
\end{equation} 
Besides the Abelian relations involving the central elements and  $\{\bar{c}_{ij}, \bar{c}_{kl} \}=0$ for disjoint pairs of indices $(ij)$ and $(kl)$, we have the defining relations:
\begin{align}
f_{ijk}&:=\frac{1}{2}\{\bar{c}_{ij},\bar{c}_{jk}\}=\frac{1}{2}\{\bar{c}_{jk},\bar{c}_{ik}\}=\frac{1}{2}\{\bar{c}_{ik},\bar{c}_{ij}\} \, \qquad (1 \leq i<j<k \leq 4) \, .
\label{fijk4r}
\end{align}
\noindent Thus the connection with the Racah algebra $R(4)$, considering that several different bases are conceivable to give a presentation (see e.g. \cite{ranktwo}), is obtained through the following identifications:
\begin{equation} 
\begin{split}
 C_i:=&-\alpha_i^2/4,\quad 1\leq i\leq 4,\\
C_{ij}:=&\bar{c}_{ij},\quad 1\leq i<j\leq 4,\\
F_{ijk}:=&f_{ijk}, \quad 1\leq i<j<k\leq 4.
\end{split}
\end{equation}
In analogy with the lower-rank case, we introduce the generators $P_{ij}$ and $P_{ii}$ (with the obvious new range for the indices) as defined in \eqref{centr}. Taking into account the symmetry properties of the generators $P_{ij}$, $F_{ijk}$, for $i \neq j \neq k \neq l \in \{1,2,3,4\}$, it is easily checked that the following relations of $R(4)$ \cite{centr} are satisfied by the elements:
\begin{equation} 
\begin{split}
 \{P_{ij}, P_{kl}\}&=0,\quad \{P_{ij}, P_{jk}\}=2 F_{ijk}, \\
 \{P_{jk}, F_{ijk}\}&=(P_{jk}+P_{jj})P_{ik}-(P_{jk}+P_{kk})P_{ij}, \\
\{P_{kl}, F_{ijk}\}&=P_{ik}P_{jl}-P_{il}P_{jk},\\
\{F_{ijk}, F_{jkl}\}&=-(F_{ijl}+F_{ikl})P_{jk} \, .
\end{split}
\end{equation}
The Hamiltonian of the model can itself be expressed as:
\begin{equation}
H=-2\sum_{1 \leq i<j}^4 P_{ij}-\sum_{i=1}^4 P_{ii}.
\label{eq:Ham}
\end{equation}

\subsection{Extrapolation to the general case $n\geq 5$}

The pattern observed for $n=4$ remains valid for values $n\geq 5$, in spite of the fact that the polynomial algebra $\mathcal{A}_n$ is of order $n-1$. From the basis (\ref{bas}) we can again consider alternative elements that reflect some symmetry or skew-symmetry properties with respect to the indices, generalizing the previous cases $f_{ijkl},g_{ijkl}$ etc to a higher number of indices. Concerning the superintegrable model on the sphere $\mathbb{S}^{n-1}$, the corresponding Hamiltonian $H$  is given by 
\begin{equation*}
H=\sum_{k=1}^{n}\frac{p_{k}^{2}}{2}+\frac{1}{2}\sum_{k=1}^{n}\frac{\alpha
_{k}^{2}}{s_{k}^{2}}=\frac{1}{2}\sum_{1\leq i<j}\left(
s_{i}p_{j}-s_{j}p_{i}\right) ^{2}+\frac{1}{2}\sum_{k=1}^{n}\frac{\alpha
_{k}^{2}}{s_{k}^{2}},
\end{equation*}
where the constraints 
\begin{equation*}
\sum_{k=1}^{n}s_{k}p_{k}=0,\;\sum_{k=1}^{n}s_{k}^{2}=1
\end{equation*}
are satisfied. The appropriate realization of $\frak{sl}\left( n\right) $
is given by 
\begin{eqnarray*}
h_{k} &=&\mathrm{i\,}\left( \alpha _{k}-\alpha _{k+1}\right) ,\;1\leq k\leq
n-1 \\
e_{ij} &=&-\frac{1}{2}\left( \left( s_{i}p_{j}-s_{j}p_{i}\right) -\mathrm{i\,%
}\left( \alpha _{i}\frac{s_{j}}{s_{i}}+\alpha _{j}\frac{s_{i}}{s_{j}}\right)
\right) \, .
\end{eqnarray*}
Considering the constants of the motion $\bar{c}_{ij}$ as defined in (\ref{redefg}), a long but routine computation leads that $H$, $\bar{c}_{ij}$ and 
$\sum_{k=1}^{n}\alpha_k^2$ are linearly dependent, as expected.  
 
\medskip
On the other hand, the Casimir invariants $c^{[k]}$ of $\mathfrak{sl}(n)$ are easily obtained as trace operators according to the formula 
\begin{equation}
c^{[k]} =\frac{1}{2}{\rm Tr} \left(
\begin{tabular}{ccccc}
$\Delta_1$ & $e_{12}$ & $\dots$ & $e_{1,n-1}$ & $e_{1n}$  \\ 
$e_{21}$ & $\Delta_2$ & $\dots$ & $e_{2,n-1}$ & $e_{2n}$  \\ 
$\vdots$ & $\vdots$ & $\dots$ & $\vdots$ & $\vdots$ \\ 
$e_{n-1,1}$ & $e_{n-1,2}$ & $\dots$ & $\Delta_{n-1}$ & $e_{n-1,n}$  \\ 
$e_{n1}$ & $e_{n2}$ & $\dots$ & $e_{n,n-1}$ & $\Delta_{n}$  \\ 
\end{tabular}
\right), 
\end{equation}
where 
\begin{equation}
\Delta_{k}= \sum_{s=k}^{n-1}\frac{n-s}{n}h_s-\sum_{s=1}^{k-1}\frac{s}{n}h_s,\quad 1\leq k\leq n.
\end{equation}
 The quadratic Casimir invariant $c^{[2]}$ is related to the Hamiltonian through the dependence relation
 \begin{equation}
2c^{[2]}+ H+\frac{n-2}{n}\left(\sum_{k=1}^n \alpha_k\right)^2=0,\quad n\geq 5,
\end{equation}
hence leading to the identity 
\begin{equation}
H=-2 c^{[2]}-\frac{n-2}{n}\left(\sum_{k=1}^{n}\alpha_k\right)^2 \, .
\end{equation}
A rather cumbersome but straightforward computation shows that for any $n\geq 5$, the Casimir invariants $c^{[k]}$ of order $k\geq 3$ collapse to a combination of $c^{[2]}$ and the constants $\alpha_i$ appearing in the Hamiltonian, hence providing the connection with the Racah algebra $R(n)$, ultimately allowing us to write the Hamiltonian as 
\begin{equation}
H=-2\sum_{1 \leq i<j}^n P_{ij}-\sum_{i=1}^n P_{ii},
\label{eq:HamN}
\end{equation}
where the generators $P_{ij}$ for $1\leq i,j\leq n$ are defined as in equation (\ref{central}). 

\section{Concluding remarks}

\noindent Using the Lie-Poisson reformulation of commutants in enveloping algebras of Lie algebras $\mathfrak{s}$, a notion of algebraic (super)integrability based on algebraic Hamiltonians and constants of the motion has been proposed. These first integrals are obtained as elements of the centralizer $C_{S(\mathfrak{s})}(\mathfrak{a})$ of a given subalgebra $\mathfrak{a}$ in the symmetric algebra of $\mathfrak{s}$. As $C_{S(\mathfrak{s})}(\mathfrak{a})$ determines a finitely-generated polynomial algebra $\mathcal{A}$, the latter can be identified with the symmetry algebra of the system defined by the algebraic Hamiltonian. For a given realization of the Lie algebra $\mathfrak{s}$, the symmetry algebra $\mathcal{A}$ can eventually reduce its polynomial order, due to additional constraints that are determined by the realization but that are not relations that hold generically in the enveloping algebra of $\mathfrak{s}$.  

\medskip
\noindent In order to illustrate the procedure, we have determined a basis for the centralizer $C_{S(\mathfrak{sl}(n))}(\mathfrak{h})$ of the Cartan subalgebra $\mathfrak{h}$ of  $\mathfrak{sl}(n)$ for any values $n\geq 2$. The resulting polynomial algebra $\mathcal{A}_n$ has been shown to be quadratic for $n=3$ and of order $n-1$ for $n\geq 4$, further defining a chain $\mathcal{A}_2\subset  \mathcal{A}_3\subset \dots \subset\mathcal{A}_n$ adapted to the canonical embedding $\mathfrak{sl}(2)\subset \mathfrak{sl}(3)\subset \dots \subset \mathfrak{sl}(n)$. It has been shown explicitly for $n=3,4$ that for suitable realizations of $\mathfrak{sl}(n)$, the polynomial algebra reduces to the $(n-2)$-rank Racah algebra $R(n)$ obtained with other techniques (see \cite{bie2018racah, bie2020racah, gab19} and references therein), providing an alternative derivation of the symmetry algebras of superintegrable models on the sphere $\mathbb{S}^{n-1}$ by means of an algebraic Hamiltonian associated to $\mathcal{A}_n$, showing in particular how the cubic algebra reduces to a quadratic one as a consequence of the realization. The construction can formally be extended to any value $n\geq 5$. This is however computationally cumbersome, as the linear dimension of the centralizer increases exponentially. So, for the values $n=5,6,7$, the dimension is given respectively by (see equation (\ref{dice}))
\begin{equation}
\dim_L C_{S(\mathfrak{sl}(5))}(\mathfrak{h})=88,\quad \dim_L C_{S(\mathfrak{sl}(6))}(\mathfrak{h})=414,\quad \dim_L C_{S(\mathfrak{sl}(7))}(\mathfrak{h})=2371,
\end{equation}
making an explicit presentation of the Poisson brackets quite complicate. In any case, the construction is feasible for any $n\geq 5$, with the centralizer $C_{S(\mathfrak{sl}(n))}(\mathfrak{h})$ determining a polynomial algebra.  The approach by commutants in higher-rank Lie algebras provides an alternative description of these superintegrable models, eventually leading to new integrable models for other appropriate realizations of the Lie algebra.  

\medskip In this context, it is natural to ask whether for the remaining classical algebras (and their real forms) the commutant of the Cartan subalgebra leads to polynomial algebras that are naturally associated to superintegrable hierarchies that have been studied by other methods. As an extension of this approach, it is also conceivable to combine the approach presented here with the so-called missing label problem \cite{Sha,C139}, where the subalgebras used have a definite physical meaning as internal symmetry algebras. Although it is expected that the computational obstructions are considerable, due to the generally complicated structure of subgroup scalars, it is not excluded that new systems with interesting properties can emerge from this ansatz. Work in this direction is currently in progress.

\section*{Acknowledgement}
RCS was   supported by
the research grant PID2019-106802GB-I00/AEI/10.13039/501100011033 (AEI/ FEDER, UE). DL and YZZ were supported by Australian Research Council Discovery Project DP190101529 (A/Prof. Y.-Z. Zhang).
IM was supported by Australian Research Council Future Fellowship FT180100099.

\addcontentsline{toc}{chapter}{bibl}
\bibliographystyle{utphys}
\bibliography{bibl}

\providecommand{\href}[2]{#2}\begingroup\raggedright\begin{thebibliography}{10}

\bibitem{perel}
A.~M. Perelomov, \href{http://dx.doi.org/10.1007/978-3-0348-9257-5}{{\em
  Integrable Systems of Classical Mechanics and Lie Algebras}}.
\newblock Birkh\"auser Verlag, Basel, 1990.

\bibitem{Tem}
P.~Tempesta, A.~V. Turbiner, and P.~Winternitz, ``Exact solvability of
  superintegrable systems''. \href{http://dx.doi.org/10.1063/1.1386927}{{\em J.
  Math. Phys.} {\bfseries 42} (2001), 4248}.

\bibitem{Trem}
{F. Tremblay, A.V. Turbiner and P. Winternitz}, ``An infinite family of
  solvable and integrable quantum systems on a plane''.
  \href{http://dx.doi.org/10.1088/1751-8113/42/24/242001}{{\em J. Phys. A.
  Math. Theor.} {\bfseries 42} (2009), 242001}.

\bibitem{Kalnins_2007}
{E. G. Kalnins, W. Miller and S. Post}, ``Wilson polynomials and the generic
  superintegrable system on the 2-sphere''.
  \href{http://dx.doi.org/10.1088/1751-8113/40/38/005}{{\em J. Phys. A: Math.
  Theor.} {\bfseries 40} (2007), 11525}.

\bibitem{Mill13}
{W. Miller, S. Post and P. Winternitz}, ``Classical and quantum
  superintegrability with applications''.
  \href{http://dx.doi.org/10.1088/1751-8113/46/42/423001}{{\em J. Phys. A:
  Math. Theor.} {\bfseries 46} (2013), 423001}.

\bibitem{Frei}
L.~Freidel and J.~M. Maillet, ``Quadratic algebras and integrable systems''.
  \href{http://dx.doi.org/10.1016/0370-2693(91)91566-E}{{\em Phys. Lett. A}
  {\bfseries 262} (1991), 278--284}.

\bibitem{Ball}
A.~Ballesteros and O.~Ragnisco, ``A systematic construction of completely
  integrable {H}amiltonians from coalgebras''.
  \href{http://dx.doi.org/10.1088/0305-4470/31/16/009}{{\em J. Phys. A: Math.
  Gen.} {\bfseries 31} (1998), 3791}.

\bibitem{Das}
C.~Daskaloyannis and Y.~Tanoudis, ``Quadratic algebras for three-dimensional
  superintegrable systems''.
  \href{http://dx.doi.org/10.1134/S106377881002002X}{{\em Phys. At. Nucl.}
  {\bfseries 73} (2010), 214–221}.

\bibitem{Jar}
L.~A. Yates and P.~D. Jarvis, ``Hidden supersymmetry and quadratic deformations
  of the space- time conformal superalgebra''.
  \href{http://dx.doi.org/10.1088/1751-8121/aab215}{{\em J. Phys. A: Math.
  Theor.} {\bfseries 51} (2018), 145203}.

\bibitem{CSIM22a}
R.~Campoamor-Stursberg and I.~Marquette, ``Quadratic algebras as commutants of
  algebraic {H}amiltonians in the enveloping algebra of {S}chr\"odinger
  algebras''. \href{http://dx.doi.org/10.1016/j.aop.2021.168694}{{\em Ann.
  Phys.} {\bfseries 437} (2022), 168694}.

\bibitem{CSIM21a}
R.~Campoamor-Stursberg and I.~Marquette, ``Hidden symmetry algebra and
  construction of polynomial algebras of superintegrable systems''.
  \href{http://dx.doi.org/10.1016/j.aop.2020.168378}{{\em Ann. Phys.}
  {\bfseries 424} (2021), 168378}.

\bibitem{Cam2022}
R.~Campoamor-Stursberg, ``On some algebraic formulations within universal
  enveloping algebras related to superintegrability''.
  \href{http://dx.doi.org/10.14311/AP.2022.62.0016}{{\em Acta Polytech.}
  {\bfseries 62} (2022), 16--22}.

\bibitem{bie2018racah}
{H. De Bie, V. X. Genest, W. van de Vijver and L. Vinet}, ``A higher rank
  {R}acah algebra and the $\mathbb{Z}_n^2$ {L}aplace-{D}unkl operator''.
  \href{http://dx.doi.org/10.1088/1751-8121/aa9756}{{\em J. Phys. A: Math.
  Theor.} {\bfseries 51} (2018), 025203}.

\bibitem{bie2020racah}
{H. De Bie, P. Iliev, W. van de Vijver and L. Vinet}, ``{The Racah algebra: An
  overview and recent results}''.
  \href{http://dx.doi.org/10.1090/conm/768/15450}{{\em Contemp. Math.}
  {\bfseries 768} (2021), 3--20}.

\bibitem{gab19}
{P. Letourneau, L. Vinet, S. Vinet and A. Zhedanov }, ``The generalized {R}acah
  algebra as a commutant''.
  \href{http://dx.doi.org/10.1088/1742-6596/1194/1/012034}{{\em J. Phys. Conf.
  Ser.} {\bfseries 1194} (2019), 012034}.

\bibitem{Correa2020}
{F. Correa, M. A. del Olmo, I. Marquette and J. Negro}, ``Polynomial algebras
  from $\mathfrak{su}(3)$ and a quadratically superintegrable model on the
  two-sphere''. \href{http://dx.doi.org/10.1088/1751-8121/abc909}{{\em J. Phys.
  A: Math. Theor.} {\bfseries 54} (2020), 015205}.

\bibitem{Lati21}
D.~Latini, I.~Marquette, and Y.-Z. Zhang, ``Embedding of the {R}acah algebra
  ${R}(n)$ and superintegrability''.
  \href{http://dx.doi.org/10.1016/j.aop.2021.168397}{{\em Ann. Phys.}
  {\bfseries 426} (2021), 168397}.

\bibitem{Dix}
J.~Dixmier, {\em Alg\`ebres enveloppantes}.
\newblock Hermann, Paris, 1974.

\bibitem{Ra}
G.~Racah, ``Sulla caratterizzazione delle rappresentazione irriducibili dei
  gruppi semisimplici di {L}ie''. {\em Rend. Acad. Naz. Lincei, Sci. Fis. Mat.
  Nat.} {\bfseries 8} (1950), 108--112.

\bibitem{Ber}
F.~A. Berezin, ``Some remarks about the associated envelope of a {L}ie
  algebra''. \href{http://dx.doi.org/10.1007/BF01076082}{{\em Funct. Anal.
  Appl.} {\bfseries 1} (1967), 91–102}.

\bibitem{Dix59}
J.~Dixmier, ``Sur l'alg\`ebre enveloppante d'une alg\`ebre de {L}ie
  nilpotente''. {\em Archiv Math.} {\bfseries 10} (1959), 321--327.

\bibitem{Alo}
I.~M. Gel'fand and A.~A. Kirillov, ``On the structure of the field of quotients
  of the enveloping algebra of a semisimple {L}ie algebra''. {\em Dokl. Akad.
  Nauk SSSR} {\bfseries 180} no~4, (1968), 775--777.

\bibitem{Bel66}
E.~Beltrametti and A.~Blasi, ``On the number of {C}asimir operators associated
  with any {L}ie group''.
  \href{http://dx.doi.org/https://doi.org/10.1016/0031-9163(66)91048-1}{{\em
  Phys. Lett.} {\bfseries 20} no~1, (1966), 62--64}.

\bibitem{RM}
R.~Campoamor-Stursberg and M.~R. de~Traubenberg,
  \href{http://dx.doi.org/doi.org/10.1142/11081}{{\em Group {T}heory in
  {P}hysics: {A} {P}ractitioner's {G}uide}}.
\newblock World Scientific, Singapore, 2018.

\bibitem{gen14}
{V. X. Genest, L. Vinet and A. Zhedanov}, ``The equitable {R}acah algebra from
  three $\mathfrak {su}(1,1)$ algebras''.
  \href{http://dx.doi.org/10.1088/1751-8113/47/2/025203}{{\em J. Phys. A: Math.
  Theor.} {\bfseries 47} (2013), 025203}.

\bibitem{lat21}
{D. Latini, I. Marquette and Y.-Z. Zhang}, ``Racah algebra {R}(n) from
  coalgebraic structures and chains of {R}(3) substructures''.
  \href{http://dx.doi.org/10.1088/1751-8121/ac1ee8}{{\em J. Phys. A: Math.
  Theor.} {\bfseries 54} (2021), 395202}.

\bibitem{Ovs}
V.~Ovsienko and A.~V. Turbiner, ``Plongements d'une alg\`ebre de {L}ie dans son
  alg\`ebre enveloppante''. {\em C. R. Acad. Sci. Paris} {\bfseries 314}
  (1992), 13--16.

\bibitem{Vin}
P.~Letourneau and L.~Vinet, ``Superintegrable systems, polynomial algebras and
  quasi-exactly solvable {H}amiltonian''.
  \href{http://dx.doi.org/10.1006/aphy.1995.1094}{{\em Ann. Phys.} {\bfseries
  243} (1995), 144--168}.

\bibitem{Mil13}
E.~G. Kalnins and W.~J. Miller, ``Quadratic algebra contractions and
  second-order superintegrable systems''.
  \href{http://dx.doi.org/10.1142/S0219530514500377}{{\em Anal. Appl.}
  {\bfseries 12} (2014), 583--612}.

\bibitem{Que}
I.~Marquette and C.~Quesne, ``Dynamical symmetry algebra of two superintegrable
  two-dimensional systems''.
  \href{http://dx.doi.org/10.1088/1751-8121/ac9164}{{\em J. Phys. A: Math.
  Theor.} {\bfseries 55} (2022), 415203}.

\bibitem{serr}
J.~P. Serre, {\em Alg\`ebres de Lie semi-simples complexes}.
\newblock W. A. Benjamin, New York, 1974.

\bibitem{Calzada_2006}
J.~A. Calzada, J.~Negro, and M.~A. del Olmo, ``Superintegrable quantum
  $\mathfrak{u}(3)$ systems and higher rank factorizations''.
  \href{http://dx.doi.org/10.1063/1.2191360}{{\em J. Math. Phys.} {\bfseries
  47} (2006), 043511}.

\bibitem{centr}
{N. Cramp\'e, J. Gaboriaud, L. Poulain d’Andecy and L. Vinet}, ``Racah
  algebras, the centralizer $\mathbb{Z}_n(\mathfrak{sl}_2)$ and its
  {H}ilbert-{P}oincar\'e series''.
  \href{http://dx.doi.org/10.1007/s00023-021-01152-y}{{\em Ann. Henri
  Poincaré} {\bfseries 23} (2022), 2657}.

\bibitem{ranktwo}
{N. Cramp\'e, L. Frappat and E. Ragoucy}, ``Representations of the rank two
  {R}acah algebra and orthogonal multivariate polynomials''.
  \href{http://dx.doi.org/10.48550/arXiv.2206.01031}{{\em arXiv:2206.01031
  [math.RT]} }.

\bibitem{Sha}
R.~T. Sharp and C.~S. Lam, ``Internal--labeling problem''.
  \href{http://dx.doi.org/10.1063/1.1664799}{{\em J. Math. Phys.} {\bfseries
  10} (1969), 2033--2037}.

\bibitem{C139}
R.~Campoamor-Stursberg, ``Internal labelling problem: An algorithmic
  procedure''. \href{http://dx.doi.org/10.1088/1751-8113/44/2/025204}{{\em J.
  Phys. A: Math. Theor.} {\bfseries 44} (2011), 025234}.

\end{thebibliography}\endgroup

\end{document}